\RequirePackage{fix-cm}
\documentclass[smallextended]{svjour3}       
\smartqed  
\usepackage{graphicx}
\usepackage{mathtools}
\usepackage{color}
\usepackage{graphicx}
\usepackage{subfig}
\usepackage{epstopdf}
\usepackage{txfonts}
\usepackage[lined,boxed,linesnumbered,commentsnumbered]{algorithm2e}
\DeclareGraphicsExtensions{.pdf,.eps,.png,.jpg,.mps}

\usepackage{array}    
\usepackage{amsmath}  
\usepackage{amssymb}
\usepackage{graphicx}
\usepackage{subfig}
\usepackage[mathscr]{eucal}
\newcommand{\mc}{\mathcal}            
\newcommand{\I}{\mathsf{I}}              

\newcommand{\G}{\mathcal{G}}   
\newcommand{\W}{\mathrm{W}}  
\newcommand{\ew}{\mathsf{w}}   

\newcommand{\ot}{\leftarrow}    

\newcommand{\conc}{\circ}             

\newcommand{\tri}{$\blacktriangleright$~}              

\newcommand*\mystrut[1]{\vrule width0pt height#1 depth0pt\relax}

\newcommand{\sjt}[1]{\textcolor{black}{#1}}
\newcommand{\plg}[1]{\textcolor{black}{#1}}

\journalname{}
\begin{document}

\title{Walk-Sums, Continued Fractions and Unique Factorisation on Digraphs}
\author{P.-L.~Giscard \and
S.~J.~Thwaite \and
D.~Jaksch
}

\institute{P.-L.~Giscard \at
             Clarendon Laboratory, Department of Physics, University of Oxford, Parks Road, Oxford OX1 3PU, United    Kingdom,\\
              Tel.: +44 1865 423263,\\
              \email{p.giscard1@physics.ox.ac.uk}
           \and
           S.~J.~Thwaite\at
              \sjt{Faculty of Physics, Ludwig Maximilian University of Munich, Theresienstrasse 37, 80333 Munich, Germany.}
              \and
              D.~Jaksch\at
              Clarendon Laboratory, Department of Physics, University of Oxford, Parks Road, Oxford OX1 3PU, United Kingdom,\\
              Centre for Quantum Technologies, National University of Singapore, 3 Science Drive 2, Singapore 117543.
}

\date{}
\maketitle
\vspace{-10mm}
\begin{abstract}
\sjt{
We show that the series of all walks between any two vertices of any (possibly weighted) directed graph $\G$ is given by a universal continued fraction of finite depth and breadth involving the simple paths and simple cycles of $\G$. A simple path is a walk forbidden to visit any vertex more than once. We obtain an explicit formula giving this continued fraction.
Our results are based on an equivalent to the fundamental theorem of arithmetic: we demonstrate that arbitrary walks on $\G$ factorize \emph{uniquely} into nesting products of simple paths and simple cycles, where nesting is a product operation between walks that we define. We show that the simple paths and simple cycles are the prime elements of the set of all walks on $\G$ equipped with the nesting product.
We give an algorithm producing the prime factorization of individual walks, and obtain a recursive formula producing the prime factorization of sets of walks. Our results have already found applications in machine learning, matrix computations and quantum mechanics.}
%
\end{abstract}
\newpage

\section{Introduction}
\subsection{Context}
Walks on graphs are pervasive mathematical objects that appear in a wide range of fields from mathematics  and physics to engineering, biology and social sciences \cite{Godsil1993,Flajolet2009,Bollobas1998,Blanchard2011,Sheskin2010,Berg1993,Borgatti2009}.  Walks are perhaps most extensively studied in the context of random walks on lattices \cite{Lawler2010}, \sjt{where they are used to model} physical processes \cite{Burioni2005}. At the same time, it is difficult to find general `context-free' results concerning \sjt{walks: indeed, the properties of walks are almost always strongly dependent on the graph on which they take place.} For this reason, many results concerning walks on graphs are \sjt{intimately connected with} the specific context in which they appear.

Over the past 30 years, the \sjt{solutions} to a number of problems across many fields have been formulated in terms of sums of walks. Amongst the most important we must mention the early work by Brydges et al. in statistical physics \cite{Brydges1983} and the seminal work by Malioutov and coworkers \cite{Malioutov2006} concerning Gaussian belief \sjt{propagation in probabilistic graphical models. These previous works are unified by two underlying themes: firstly, that some quantities are most naturally expressed as sums of walks, and secondly, that these walk-sums can be reduced to more manageable expressions by resumming certain families of terms appearing in the sum. However, none of the existing studies address the question of how these resummations can be developed in a systematic fashion. Consequently, the results in the existing literature depend strongly on the context of their discovery, and are only applicable in a limited number of situations. The general feasibility of walk resummations for graphs of arbitrary structure thus remains an open problem. In this article we present a mathematically rigorous and general approach to the question of summing and resumming walks. In particular, we obtain an explicit expression for the sum of all walks between any two vertices of any weighted multi-digraph.} \plg{We have already produced applications for our results in the fields of machine learning \cite{Giscard2014b}, matrix computations \cite{Giscard2011b,Giscard2014a} and quantum dynamics \cite{Giscard2014}.}

\subsection{\sjt{A systematic approach to walk sums}}
\sjt{In this work we consider walks on (possibly weighted) directed graphs as mathematical entities in their own right.} 
We demonstrate that \sjt{these walks exhibit non-trivial properties that are largely \emph{independent} of the digraph on which they} take place. Foremost amongst these properties is \sjt{that any walk can be uniquely factorized into a product of prime walks, which we show are precisely the simple paths and simple cycles of the underlying graph.}\footnote{Simple paths and simple cycles are also known as self-avoiding walks and self-avoiding polygons, respectively.}. \sjt{An important consequence of this result is the existence of a universal closed-form expression} for the series of all walks between any two vertices of any finite (weighted) digraph: \sjt{namely,} a branched continued fraction of finite depth and breadth, which we provide. \sjt{This continued fraction} is the prime representation of the walk series, \sjt{an analog of the Euler product formulae} for the Riemann zeta function and other totally multiplicative functions in number theory. This universal continued fraction, which we present and prove here, has already found applications in the fields of matrix functions \cite{Giscard2011b}, differential calculus \cite{Giscard2014a}, quantum dynamics \cite{Giscard2014} and machine learning \cite{Giscard2014b}. Although seemingly disparate, many open questions in these disciplines are unified by their natural formulation in terms of sums of walks and thus benefit from the results presented here.

\sjt{The usual product operation} on the set $W_\G$ of all walks on a digraph $\G$ \sjt{is concatenation, which we denote here by $\conc$.}
\sjt{Concatenation} is a very liberal operation: the concatenation $a\conc b$ of \sjt{two walks $a$ and $b$} is
non-zero whenever the final vertex of $a$ is the same as the initial vertex of $b$. This implies that both the
irreducible and the prime elements of the set of all walks equipped with the concatenation product, denoted $(W_\G,\conc)$, are the walks of \sjt{length 1 on $\G$: in other words, the edges} of $\G$.
Consequently, the factorisation of a walk $w$ on $\G$ into concatenations of prime walks is somewhat trivial. For this reason, we abandon the operation of concatenation and \sjt{define instead a new product between walks, which we term nesting and denote by $\odot$.} Nesting is a much more restrictive operation than concatenation, in that the nesting of two walks is non-zero only if the walks satisfy certain constraints.
As a result of these constraints, the \sjt{irreducible and prime elements of $(W_\G,\,\odot)$ are the simple paths and simple cycles of $\G$, rather than the edges of $\G$.} The rich structure that \sjt{the nesting operation induces} on walk sets is at the origin of the universal continued fraction formula for formal series of walks.
%
%
%

This article is organised as follows. In \S \ref{concepts}, we present the notation and terminology used throughout the article. In particular, we define the nesting product and establish its properties in  \S \ref{NestingDef}.
In \S \ref{sec:WalkFactor} we obtain \sjt{our central result: we prove the existence and uniqueness of the factorization of any walk on a digraph $\G$ into nesting products of primes (i.e.~the simple paths and simple cycles on $\G$).} We provide an \sjt{algorithm that produces} the prime factorisation of individual walks in \S\ref{Algosection} and a recursive formula to reduce sets of walks into nested sets of primes in \S\ref{SectionFactorBasis}. \sjt{In \S\ref{PathSumPrimeSeries} we exploit these results to present the prime representation of walk series.} \sjt{Specifically, we obtain in \S\ref{formal} an explicit branched continued fraction for the sum of all walks between any two vertices of any digraph, and in \S\ref{weightedsum} extend this result to the case of weighted digraphs.} Finally, the last section \S\ref{StarHeightsection} is devoted to \sjt{identifying the maximum} depth of this continued fraction.

\section{Required Concepts}\label{concepts}
\subsection{Notation and terminology}\label{NotationCh1}
A \emph{directed graph} or \emph{digraph} is a set of \emph{vertices} connected by \sjt{\emph{directed edges},} also known as arrows. An arrow $e$ starts at vertex $s(e)$ and terminates at vertex $t(e)$, which we write $e:\,s(e)\rightarrow t(e)$ or $(s(e)t(e))$.
Throughout this article, we let $\G=\big(\mc{V}(\G),\mc{E}(\G)\big)$ be a finite digraph with $\mc{V}(\mc{G})$ its vertex set and $\mc{E}(\mc{G})$ its edge set. This digraph may contain self-loops but not multiple edges, i.e.~we restrict ourselves to at most one directed edge from $\alpha\in\mc{V}(\G)$ to $\omega\in\mc{V}(\G)$. The latter restriction is solely for the purpose of notational clarity, and all of our results can be straightforwardly extended to cases where $\mc{G}$ contains multiple edges. We denote the vertices of $\G$ by numbers or Greek letters \sjt{$\alpha,\beta,\ldots$, as convenient.} The digraph obtained by deleting vertices $\alpha,\beta,\ldots$ and all edges incident on these vertices from $\G$ is written $\G\backslash\{\alpha,\beta,\ldots\}$.

A \emph{walk} $w$ of length $\ell(w)=n\ge1$ from $\mu_0$ to $\mu_{n}$ on $\G$ is a left-to-right sequence \sjt{$(\mu_0\mu_1),(\mu_1\mu_2),\cdots,(\mu_{n-1}\mu_n)$} of $n$ contiguous directed edges. This walk starts at $\mu_0$ and terminates at $\mu_n$
We \sjt{represent} $w$ by its vertex string $\mu_0\,\mu_1\,\mu_2\,\cdots\,\mu_n$ or by its vertex-edge sequence $(\mu_0)(\mu_0\mu_1)(\mu_1)\cdots\,(\mu_{n-1}\mu_n)(\mu_n)$.
If $\mu_0=\mu_n$, $w$ is termed a \textit{cycle} or closed walk; otherwise, $w$ is an open walk. The initial and final vertices of $w$ are called its \textit{head} and \textit{tail}, respectively. When necessary, they will be denoted $h(w)$ and $t(w)$. 
The set of all walks on $\G$ is denoted by $W_{\G}$, and the set of all walks from vertex $\mu_0$ to vertex $\mu_n$ on $\G$ is denoted by $W_{\G;\,\mu_0\mu_n}$.

A \emph{simple path} is an open walk whose vertices are all distinct. The set of all the simple paths on $\G$ is denoted by $\Pi_{\G}$. The set of simple paths from $\alpha$ to $\omega$ is denoted by $\Pi_{\G;\,\alpha\omega}$. On any finite digraph $\G$, these sets are finite.

A \emph{simple cycle} is a cycle whose internal vertices are all distinct and different from the initial vertex. The set of all the simple cycles on $\G$ is denoted by $\Gamma_{\G}$, while the set of simple cycles off a specific vertex $\alpha$ is denoted by $\Gamma_{\G;\,\alpha}$. On any finite digraph $\G$, these sets are finite.

A \emph{trivial walk} is a walk of length 0 off any vertex $\mu \in \mathcal{V}(\G)$, denoted by $(\mu)$. \sjt{A trivial walk is a simple path, but not a simple cycle. Note that trivial walks are different from the \emph{empty walk}, denoted $0$, whose length is undefined.}

The \emph{concatenation} is a non-commutative product operation between walks. Let $w_1=\alpha_0\cdots\alpha_{\ell}\in W_\G$ and $w_2=\beta_0\cdots\beta_{\ell'}\in W_\G$. Then the concatenation of $w_1$ with $w_2$ is defined as
\begin{equation}
w_1\conc w_2=
\begin{cases}
\alpha_0\cdots\alpha_\ell\,\beta_1\cdots\beta_{\ell'},&\textrm{if }\alpha_\ell\equiv \beta_0,\\
0,&\textrm{otherwise}.
\end{cases}
\end{equation}
The empty walk is absorbing for the concatenation, \sjt{i.e. for every $w\in W_{\G}$, we have $w\conc 0=0\conc w=0$.}
%
%

\subsection{The nesting product}\label{NestingDef}
We now turn to the definition and properties of the nesting product. Nesting is more restrictive than concatenation; in particular, \sjt{the nesting product of two walks is non-zero} only if they obey the following property:

\begin{definition}[Nestable property]\label{CanDef}
Consider two walks $(w_1,w_2)\in W_{\G}^2$ with $w_2=\beta\,\beta_1\,\cdots\, \beta_{\ell_2-1}\, \beta$ a cycle from $\beta$ to itself, and $w_1=\alpha_0\,\alpha_1\,\cdots\beta\cdots\, \alpha_{\ell_1}$ a walk that visits $\beta$ at least once. \sjt{Let $\alpha_j=\beta$ be the first appearance of $\beta$ in $w_1$.} Then the couple $(w_1,w_2)$ is nestable if and only if one of the following conditions holds:
\begin{itemize}
\item[(i)] $w_1$ and $w_2$ are cycles off the same vertex $\beta$; or
\item[(ii)] no vertex that $w_1$ visits before reaching $\beta$ for the first time is also visited by $w_2$. 
\end{itemize}
\end{definition}

The nestable property describes the natural structure arising from the cycle-erasing  
procedure \sjt{(also known as loop-erasing procedure) introduced by Lawler in 
\cite{Lawler1980,Lawler2010}. Consider traversing a walk $w$ on a graph, removing 
all simple cycles $c_i$ (where $1\leq i\leq n$) from $w$ in chronological order. Upon 
reaching the end of the walk, the surviving vertex string forms a simple path $p$. It 
can be seen that for any eliminated cycle $c$ whose head is on $p$, the couple $
(p,c)$ is nestable.} \plg{Similarly, if the head of a cycle $c_i$ is an internal vertex of another cycle $c_j
$, where $j>i$, then the couple $(c_j,c_i)$ is nestable.} This provides a natural motivation for the nesting product as the inverse operation of the cycle-erasing procedure: the walk $w$ can be written as a nesting product involving the simple path $p$ and the erased simple cycles $c_1\cdots c_n$.

\begin{definition}[Nesting product]\label{def:Nesting}
Let $(w_1,\,w_2)\in W_\G^2$ be two walks on $\G$. If the couple $(w_1,\,w_2)$ is not nestable, we define the nesting product to be $w_1\odot w_2=0$. Otherwise, let $w_1 = \alpha_0\,\alpha_1\cdots \beta\cdots \alpha_{\ell_1}$ be a walk of length $\ell_1$ and let $w_2 =\beta\,\beta_1\cdots\beta_{\ell_2-1}\,\beta$ be a cycle of length $\ell_2$ from $\beta$ to itself. Then the operation of nesting is defined by
\begin{subequations}
\begin{eqnarray}
\hspace{-5mm}\odot:~W_{\G}\times W_{\G}&\to& W_{\G},\\
(w_1,\,w_2)&\to&w_1\odot w_2=\alpha_0\,\alpha_1\cdots\beta\,\beta_1\cdots\beta_{\ell_2-1}\,\beta\cdots \alpha_{\ell_1}.
\end{eqnarray}
\end{subequations}
The walk $w_1\odot w_2$ of length $\ell_1+\ell_2$ is said to consist of $w_2$ nested into $w_1$. The vertex sequence of $w_1\odot w_2$ is formed by replacing the \plg{\emph{last}} appearance of $\beta$ in $w_1$ by the entire vertex sequence of $w_2$.
\end{definition}

\begin{figure}[t!]
\begin{center}
\vspace{2mm}
\includegraphics[width=1 \textwidth]{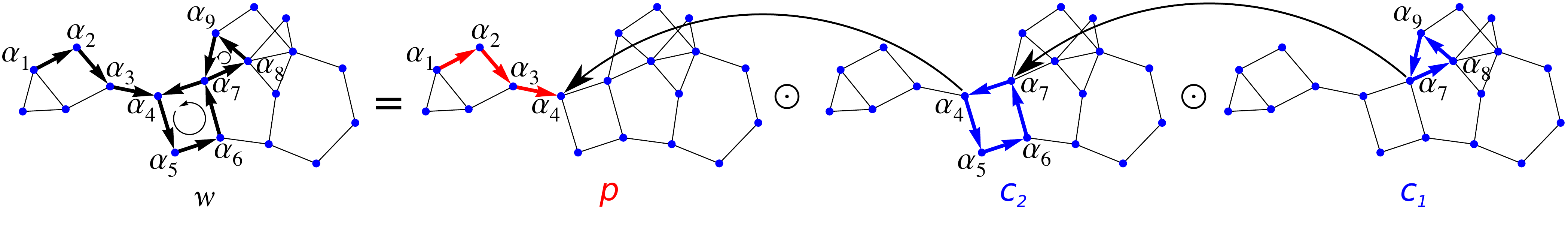}
\end{center}
\vspace{-2mm}
\caption[An example of nesting]{An example of nesting: the walk $w=\alpha_1\alpha_2\alpha_3\alpha_4\alpha_5\alpha_6\alpha_7\alpha_8\alpha_9\alpha_7\alpha_4$ is obtained upon inserting the triangle $c_1=\alpha_7\alpha_8\alpha_9\alpha_7$ into the square $c_2=\alpha_4\alpha_5\alpha_6\alpha_7\alpha_4$ and then into the simple path $p=\alpha_1\alpha_2\alpha_3\alpha_4$, that is $w=p\odot\big(c_2\odot c_1\big)$.}
\end{figure}

\noindent\tri Nesting is non-commutative and non-associative: for example, $11\odot131=1131$, while $131\odot11=1311$, and $\big(12\odot242\big)\odot11=11242$, while $12\odot\big(242\odot11\big)=0$.

\vspace{1.5mm}
\noindent\tri Nesting coincides with concatenation for cycles off the same vertex: for $(c_1,c_2)\in W_{\G;\,\alpha\alpha}^2$ we have $c_1\odot c_2=c_1\conc c_2$. Consequently, nesting is associative over the cycles: if $c_3\in W_{\G;\,\alpha\alpha}$ then $(c_1\odot c_2)\odot c_3=c_1\odot (c_2\odot c_3)=c_1\odot c_2\odot c_3$. This in turn implies power-associativity over the cycles, so we can simply write $c^p$ for the nesting of a cycle $c$ with itself $p$ times, e.g. $1212121=121\odot121\odot121=121^3$. We interpret $c^0$ as the trivial walk off $h(c)$.

\vspace{1.5mm}
\noindent\tri \plg{Let $\mu\in\mathcal{V}(\G)$. Consider the trivial walk $(\mu)$ and observe that for any cycle $w \in W_{\G;\,\mu\mu}$ from $\mu$ to itself we have $(\mu)\odot w=w$. Therefore we say that the trivial walk $(\mu)$ is a left-identity element on the cycles from $\mu$ to itself.} 

\vspace{2mm}Finally, with the nesting product comes a notion of divisibility. This notion plays a fundamental role in the identification of irreducible and prime walks:
\begin{definition}[Divisibility]\label{def:divisibility}
Let $w$ and $w'$ be two walks. We say that $w'$ divides $w$, and write $w'\,|\,w$, if and only if $w$ can be written using non-zero nesting products involving $w'$.
A walk $w'$ that divides $w$ will be called a factor or divisor of $w$.
%
\end{definition}

\noindent This basic definition is equivalent to the following, more explicit formulation:
\begin{definition}
Let $w$ and $w'$ be two walks. Then we say that $w'$ divides
$w$, and write $w' \,|\,w$, if and only if one of the following conditions holds:
\begin{itemize}
\vspace{-1mm}
\item[(i)] there exists a walk $\plg{w''\neq w}$ such that $w'|w''$ and $w''|w$; or
\item[(ii)] there exists $n\geq 0$ walks $w_1,\,w_2,\,\cdots,\, w_n$  and an integer $0\leq i\leq n$ such that $w=w_1\odot\cdots\odot w_i\odot w'\odot w_{i+1}\odot\cdots\odot w_n$.
\end{itemize}
\end{definition}

\section{Prime Factorisation on Digraphs}\label{sec:WalkFactor}
In this section, we prove the existence and uniqueness of the factorisation of individual walks on digraphs into nesting products of prime walks, which we identify to be the simple paths and simple cycles. We provide an algorithm \sjt{that factors} walks.
Second, we give an explicit formula expressing the set of all walks between any two vertices of any digraph \sjt{as a collection of nested sets} of prime walks. \sjt{We will use this result in the next section} to obtain representations for the series of all walks on any digraph which involve only prime walks. For each result we present, we provide a simple example demonstrating its use.

\subsection{Existence and uniqueness of the prime factorisation of walks}\label{FTAdigraph}
The fundamental theorem of arithmetic is arguably the most important \sjt{result in number theory} \cite{Hardy1979}. It establishes the central role \sjt{played by the prime numbers, and has many profound consequences on} the properties of integers. We now present its analogue for individual walks on arbitrary digraphs.

We begin by presenting the notion \sjt{of a factorisation of a walk and stating the conditions under which two factorisations are equivalent.}
%
\sjt{A factorisation of a walk $w$ on $\G$, denoted $\mathsf{Fac}\,w$, is a decomposition of $w$ into a nesting product of other walks on $\G$, which we term the factors of $w$. Recombining the factors of $w$ with the nesting operation reproduces the original walk $w$.} Since nesting is neither commutative nor associative, it is necessary to be cautious about equivalent factorisations of a walk. We say that two factorisations $\mathsf{Fac}\,w$ and $\mathsf{Fac}'\,w$ of a walk $w$ are equivalent, denoted $\mathsf{Fac}\,w\equiv \mathsf{Fac}'\,w$, if and only if one can be obtained from the other through the reordering of parentheses and factors, and up to nesting with trivial walks, without modifying $w$. In particular, equivalent factorisations of a walk are made up of the same non-trivial factors. Equivalent factorisations are generated by the following operations:
\begin{itemize}
\vspace{-1mm}
\item[i)] Multiplication by trivial factors (i.e. trivial walks).
\item[ii)] Let $w, a$ and $b$ be walks such that $(w,a)$ and $(w,b)$ are nestable. Then one can replace $(w\odot a)\odot b$ by $(w\odot b)\odot a$ if and only if $a$ and $b$ do not have any vertex in common.
\item[iii)] Let $w, a$ and $b$ be walks such that $(w,a)$ and $(a,b)$ are nestable but $(w,b)$ is \emph{not} nestable. Then one can replace $(w\odot a)\odot b$ by $w\odot (a \odot b)$.
\end{itemize}
\vspace{-1mm}From now on, we shall speak of walk factorisations up to equivalence, that is up to the application \sjt{of one or more} of the above operations.

\vspace{2mm}Of particular interest is the factorisation of a walk into nesting products of prime walks, called the \textit{prime factorization}. Following standard definitions \cite{Lang2002,Lam2001}, a walk $w$ is said to be \textit{prime} with respect to nesting if and only if for all nestable couples of walks $(w',w'')$ such that $w\,|\,(w'\odot w'')$ then $w\,|\,w'$ or $w\,|\,w''$. It is a central result of this article that the prime factorisation of any walk $w$ exists and is unique and the set of primes factors of $w$ is uniquely determined by $w$:

\begin{theorem}[Prime factorisation of walks]\label{FundamentalthmDigraph}
Any walk on $\G$ factorises uniquely into nesting products \sjt{of prime walks:} the simple paths and simple cycles on $\G$.
\end{theorem}
The theorem makes three statements concerning the prime factorisation of a walk: i) it always exists; ii) it is unique; and iii) a walk is prime if and only if it is a simple path or a simple cycle.



\vspace{2mm}The proof of Theorem \ref{FundamentalthmDigraph} is organized as follows. We begin by showing that a walk is \sjt{irreducible -- that is, that it cannot be expressed as a nesting product of two or more non-trivial walks -- if and only if} it is a simple path or a simple cycle. Second, we show that the factorisation of a walk into nesting products of irreducible walks always exists and is unique. Third, we prove that a walk is prime if and only if it is a simple path or a simple cycle. \sjt{Taken together, these steps} establish Theorem \ref{FundamentalthmDigraph} as an equivalent to the fundamental theorem of arithmetic.\\

\textit{Proof }
Following standard definitions \cite{Lang2002,Lam2001}, a walk $w$ is \textit{irreducible} if, whenever there exists a divisor $w'$ of $w$, then either $w'$ is trivial, or $w'=w$ up to nesting with trivial walks (i.e. local identities). \sjt{A walk that is not irreducible is said to be \textit{reducible}. Then we have the following result:}

\begin{lemma}\label{irreducible}
A walk $w$ is irreducible if and only if it is a simple path or a simple cycle.
\end{lemma}
\begin{proof}
The backward direction is straightforward since simple paths and simple cycles have no repeated internal vertices and are thus irreducible.
For the forward direction, consider an irreducible walk $w$ and suppose that $w$ is neither a simple path nor a simple cycle. We distinguish two cases: i) if $w$ is an open walk or $w$ is closed and does not have its head as an internal vertex, then there exists an earliest vertex $\mu$ visited at least twice by $w$ (earliest internal vertex $\mu$ if $w$ is closed). Then let $s_\mu\subset w$ be the vertex sequence joining the first appearance of $\mu$ to its final appearance in $w$, and let $w_\mu$ be the vertex sequence obtained from $w$ by replacing $s_\mu$ by $(\mu)$ in $w$. Then $(w_\mu,s_\mu)$ is a nestable couple of non-trivial walks and $w=w_\mu\odot s_\mu$, which is a contradiction. ii) If instead $w$ is a closed walk and its head $\mu:=h(w)$ also appears as an internal vertex, then let $s_{\mu}$ be the vertex sequence joining the second appearance of $\mu$ to its last appearance in $w$ and let $w_{\mu}$ be the vertex sequence obtained from $w$ by replacing $s_{\mu}$ by $(\mu)$ in $w$. Then again $(w_{\mu},s_{\mu})$ is a nestable couple of non-trivial walks and $w=w_{\mu}\odot s_{\mu}$, which is a contradiction.\qed
\end{proof}


\begin{lemma}\label{factirred}
Any walk on $\G$ factorises uniquely into \sjt{nesting products of irreducible} walks.
\end{lemma}

\begin{proof}
We prove the lemma by induction on the walk length. Let $\mathcal{P}(n)$ be the following proposition: \textit{for any walk $w$ of length $\ell(w)\leq n$, there exists a unique factorisation of $w$ into nesting products of irreducible walks, denoted $\mathsf{Fac}_I\,w$.}\\

\noindent\textit{Base case:} we establish $\mathcal{P}(1)$. Consider a walk $w$ of length $\ell(w)=1$. Then $w$ is either a self-loop $\alpha\alpha$ or comprises a single edge $\alpha\omega$, for some vertices $\alpha$ and $\omega$. \sjt{In either case} it is irreducible. Furthermore, the factorised form of $w$ is $w$ itself and is clearly unique, so $\mathcal{P}(1)$ holds.\\

\vspace{-1mm}\noindent\textit{Induction:} \sjt{we show that for any $n \ge 1$, the ensemble of statements $\mathcal{P}(1),\mathcal{P}(2),\ldots,\mathcal{P}(n)$ imply $\mathcal{P}(n+1)$.} To this end, consider a walk $w$ of length $\ell(w)=n+1$. If $w$ is irreducible, then its \sjt{factorisation exists and is unique: this factorisation is $w$ itself, $\mathsf{Fac}_I\,w\equiv w$.}

In the case where $w$ is reducible, we begin by proving that \sjt{it has at least} one factorisation into products of irreducible walks. If $w$ is reducible, then there exists a nestable couple $(a,b)$ of non-trivial walks such that $w=a\odot b$. Necessarily $1\leq\ell(a),\,\ell(b)\leq n$ \sjt{and so by the induction hypothesis there exist unique factorisations of $a$ and $b$} into products of irreducible walks, $\mathsf{Fac}_I\,a$ and $\mathsf{Fac}_I\,b$. Then $\mathsf{Fac}_I\,w:=\mathsf{Fac}_I\,a\odot \mathsf{Fac}_I\,b$ is a valid factorisation of $w$ into irreducible walks.

Now suppose that there exists a second factorisation $\mathsf{Fac}'_I\,w$ of $w$ into irreducibles, which is \sjt{different from the first:} $\mathsf{Fac}_I\,w\nequiv \mathsf{Fac}_I'\,w$. Since $\mathsf{Fac}_I'\,w$ exists, there exists a nestable couple $(c,d)$ of non-trivial \sjt{walks such that $w=c\odot d$, $1\leq\ell(c)\le n$, $1 \le \ell(d)\leq n$, and $\mathsf{Fac}_I'\,w=\mathsf{Fac}_I\,c\odot \mathsf{Fac}_I\,d$.}

\sjt{Consider the case where $a=c$ and $b=d$. Then since $a$ and $b$ have length not greater than $n$, it follows from the induction hypothesis that the factorisation of $a$ into products of irreducible walks exists and is unique. Thus $a=c$ implies that $a$ and $c$ have the same factorisation into irreducible walks: $\mathsf{Fac}_I\,a\equiv \mathsf{Fac}_I\,c$. By analogous reasoning, $b=d$ implies that $\mathsf{Fac}_I\,b\equiv \mathsf{Fac}_I\,d$. In this situation $\mathsf{Fac}_I\,w$ and $\mathsf{Fac}_I'\,w$ are equivalent and $\mathcal{P}(n+1)$ holds.}

Otherwise, consider the case where $(a,b)\neq (c,d)$. Since $w=a\odot b=c\odot d$, the position of the cycle $d$ in the vertex sequence of $w$ must fall into one of the following three cases:
\begin{itemize}
\item[i)] \sjt{\underline{$d$ is included in $a$: $d\subseteq a$}}. Let $e:=a\cap c$ be the vertex sequence common to $a$ and $c$ \sjt{(see schematic representation below).}
Since $(a,b)$ is nestable $(e,b)$ is nestable, and similarly $(c,d)$ nestable implies $(e,d)$ nestable. Additionally $c=e\odot b$ and $a=e\odot d$. It follows that $\mathsf{Fac}_I\,w=\mathsf{Fac}_I\,a\odot \mathsf{Fac}_I\,b=\big(\mathsf{Fac}_I\,e\odot \mathsf{Fac}_I\,d\big)\odot \mathsf{Fac}_I\,b$ and $\mathsf{Fac}_I'\,w=\mathsf{Fac}_I\,c\odot \mathsf{Fac}_I\,d=\big(\mathsf{Fac}_I\,e\odot \mathsf{Fac}_I\,b\big)\odot \mathsf{Fac}_I\,d$. \sjt{Now since each of $e$, $b$ and $d$ have} length less than or equal to $n$, by the induction hypothesis their factorisations into irreducible walks exist and are unique. Consequently $\mathsf{Fac}_I\,w$ and $\mathsf{Fac}_I'\,w$ are equivalent and $\mathcal{P}(n+1)$ holds.
\begin{figure}[h!]
\centering
\includegraphics[scale=.45]{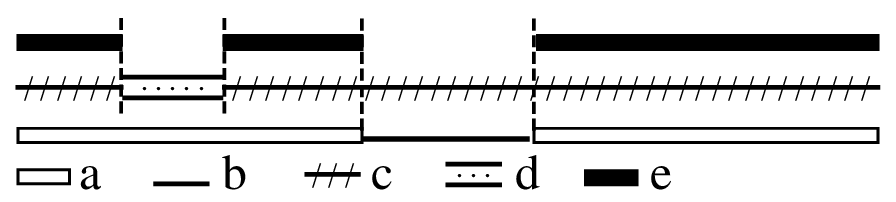}
\caption*{\small Schematic representation of the vertex sequence of $w$ in the case $d\subseteq a$.} \label{fig:case1}
\end{figure}
\newpage
\item[ii)]\underline{$d$ is included in $b$, $d\subseteq b$}. We proceed similarly \sjt{to case i).} Let $e:=b\cap c$ be the vertex sequence common to $b$ and $c$. \sjt{By construction, $e$ is a cycle. The couple $(a,b)$ and therefore $(a,e)$ are nestable, while the fact that $(c,d)$ is nestable implies that $(e,d)$ is also nestable. Thus we have $b=e\odot d$ and $c=a\odot e$, while $\mathsf{Fac}_I\,w=\mathsf{Fac}_I\,a\odot \mathsf{Fac}_I\,b=\mathsf{Fac}_I\,a\odot \big(\mathsf{Fac}_I\,e\odot \mathsf{Fac}_I\,d\big)$ and $\mathsf{Fac}_I'\,w=\mathsf{Fac}_I\,c\odot \mathsf{Fac}_I\,d=\big(\mathsf{Fac}_I\,a\odot \mathsf{Fac}_I\,e\big)\odot \mathsf{Fac}_I\,d$. Now since each of $a$, $e$ and $d$ has length less than or equal to $n$, by the induction hypothesis their factorizations into irreducible walks exist and are unique.} Consequently $\mathsf{Fac}_I\,w$ and $\mathsf{Fac}_I'\,w$ are equivalent and $\mathcal{P}(n+1)$ holds.
\begin{figure}[h!]
\centering
\includegraphics[scale=.45]{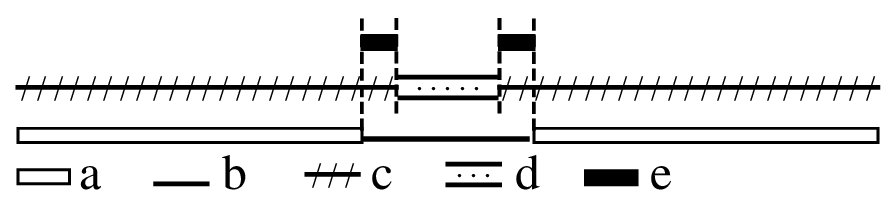}
\caption*{\small Case $d\subseteq b$.} \label{fig:case2}
\vspace{-4mm}
\end{figure}
\item[iii)]\underline{$d$ straddles $a$ and $b$}. This case is essentially different from \sjt{i) and ii), and it is necessary to} distinguish subcases. Given that $a\odot b\neq0$ and $c\odot d\neq0$ by assumption, $b$ and $d$ must \sjt{both be cycles. Let the head vertex of $b$ be $\beta$, and the head vertex of $d$ be $\delta$.} Since $d$ straddles over $a$ and $b$, then $\delta$ is visited by both $a$ and $b$, and $\beta$ is visited by both $c$ and $d$.
\begin{figure}[h!]
\centering
\includegraphics[scale=.45]{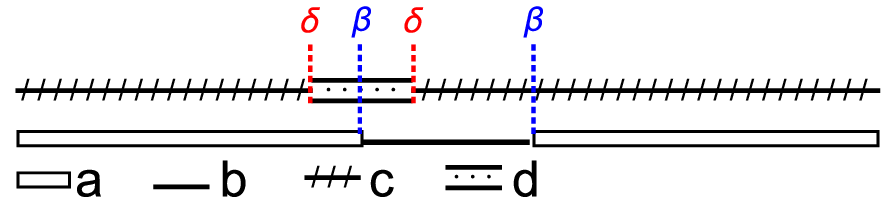}
\caption*{\small Case $d$ straddles over $a$ and $b$.} \label{fig:case3}
\vspace{-4mm}
\end{figure}

We first examine the situation where $\beta$ and $\delta$ are \sjt{different vertices. Then two cases exist:}\\
\textit{1) $a$ is a cycle off $\beta$}. Then $c$ visits $\beta$ before $\delta$, but $\beta$ is also visited by $d$. Thus the couple $(c,d)$ is non-nestable, which is a contradiction.\\
\textit{2) $a$ is an open walk from $\alpha$ to $\beta$, or $\beta$ appears only as an internal vertex of $a$}. In either situation, $a$ visits $\delta$ before the final appearance of $\beta$, but $\delta$ is also visited by $b$. Then the couple $(a,b)$ is non-nestable, which is a contradiction.

Finally, it remains to consider the case where $\beta$ and $\delta$ are identical. Then the last appearance of $\beta$ in $c$ must be the last vertex of $b$ (since $a\odot b$ nests $b$ into $a$ off the final appearance of $\beta$). However, $d$ straddles $a$ and $b$, which implies that $c$ visits $\beta$ after the last vertex of $d$; and $d$ is not nested into $c$ off the last appearance of $\beta$, a contradiction.  \sjt{Therefore case iii) is impossible:}  $w=a\odot b=c\odot d$ cannot hold with $d$ straddling $a$ and $b$.

\end{itemize}

\vspace{1mm}\noindent We have demonstrated that $\mathcal{P}(1)$ is true, and upon supposing that $\mathcal{P}(j)$ holds for all $j\leq n$, we have shown that $\mathcal{P}(n+1)$ holds. Consequently $\mathcal{P}(n)$ \sjt{holds for all $n \ge 1$.} The factorisation of a walk into nesting products of irreducible walks thus always exists and is unique.\qed
\end{proof}

%

We complete the proof of Theorem \ref{FundamentalthmDigraph} by establishing that simple paths and simple cycles are prime:
\begin{lemma}\label{PrimeProof}
Let $w$ be a walk. Then $w$ is prime \sjt{if and only if it is} a simple path or a simple cycle.
\end{lemma}
\begin{proof}
Firstly, we prove the backward direction: \sjt{that if $w$ is a simple path or a simple cycle, then $w$ is prime.} Consider a walk $w$ and a nestable couple $(w_1,w_2)$ such that $w\,|\,(w_1\odot w_2)$. Since $w$ \sjt{is either a simple path or a simple cycle,} then, by Lemma \ref{irreducible}, $w$ is an irreducible factor appearing in the factorisation $\mathsf{Fac}_I\,(w_1\odot w_2)$ of $w_1\odot w_2$ into nesting products of irreducible walks. By uniqueness of this factorisation, Lemma \ref{factirred}, $w$ must either be an irreducible factor of $\mathsf{Fac}_I\,w_1$, implying $w\,|\,w_1$; or an irreducible factor of $\mathsf{Fac}_I\,w_2$, implying $w\,|\,w_2$; or both. It follows that $w$ is prime.
Secondly, we prove the forward direction: $w$ prime $\Rightarrow w$ is a simple path or a simple cycle. Suppose that there exists a prime walk $w$ which is neither a simple path nor a simple cycle. Then by, Lemma \ref{irreducible}, $w$ is reducible and there exists at least one nestable couple $(w_1,w_2)$ of non-trivial walks such that $w=w_1\odot w_2$. Clearly $w\,|\, (w_1\odot w_2)$, and $w_1$ and $w_2$ are strictly shorter than $w$. Therefore $w$ divides neither $w_1$ nor $w_2$, which is a contradiction.\qed
\end{proof}

\noindent \sjt{Taken together,} Lemmas \ref{irreducible}, \ref{factirred} and \ref{PrimeProof} establish Theorem \ref{FundamentalthmDigraph}.\qed

\subsection{An algorithm to factorise individual walks}\label{Algosection}
\sjt{Let $w$ be a walk on a digraph $\G$, and $S_{\!R}\big(\mathsf{Fac}\,w\big)$ be the set of \emph{reducible} factors appearing in a factorisation $\mathsf{Fac}\,w$ of $w$; that is
\begin{equation*}
  S_{\!R}\big(\mathsf{Fac}\,w\big) = \big\{ w' : w' \in \mathsf{Fac}\,w \text{ and $w'$ is reducible}  \big\}.
\end{equation*}
Algorithm} \ref{AlgoCanFac}, presented on p. \pageref{AlgoCanFac}, \sjt{then proceeds as follows. We begin by setting $\mathsf{Fac}\,w:=w$. Then an arbitrary reducible factor $a$ of $\mathsf{Fac}\,w$ is chosen and factorized into a nesting product of strictly shorter walks, yielding a factorization $\mathsf{Fac}\,a$. Next, $\mathsf{Fac}\,w$ is updated by replacing $a$ by its factorization $\mathsf{Fac}\,a$, an operation which we denote $\mathsf{Fac}\,w\rightarrow \mathsf{Fac}\,w\,/\,\{a\rightarrow \mathsf{Fac}\,a\}$. Finally, another reducible factor appearing in the updated factorisation $\mathsf{Fac}\,w$ is chosen, and the process is repeated. At each round, reducible factors are decomposed into nesting products of shorter walks. The algorithm stops when $S_{\!R}\big(\mathsf{Fac}\,w\big)$ is the empty set $\varnothing$, at which point $\mathsf{Fac}\,w$ is the prime factorisation of $w$.}

\setlength{\textfloatsep}{15pt}
\begin{algorithm}[t!]
\SetKwInOut{Input}{Input}\SetKwInOut{Output}{Output\,}
\Input{A walk $w\in W_\G$}
\Output{The prime factorisation of $w$}
\BlankLine
$\mathsf{Fac}\,w:= w$
\vspace{1mm}
\While{\label{While1}$S_{\!R}\big(\mathsf{Fac}\,w\big)\neq\varnothing$}{
Choose any $a\in S_{\!R}\big(\mathsf{Fac}\,w\big)$
\vspace{1mm}
\eIf{$h(a)=t(a)=\mu$ and $a$ visits vertex $\mu$ a total of $k>2$ times}{
Let $c_{1},\cdots, c_{k-1}$ be the $k-1$ cycles off $\mu$ identified by splitting the vertex string of $a$ at each internal appearance of $\mu$.\\
\vspace{.5mm}
$\mathsf{Fac}\,w\rightarrow \mathsf{Fac}\,w\,/\,\{a\rightarrow(c_{1}\odot\cdots\odot c_{k-1})\}$
\hspace{5mm}\CommentSty{\% Replace $a$ with $(c_{1}\odot\cdots\odot c_{k-1})$ in $\mathsf{Fac}\,w$}\\}{
$w_0:=a$; $\mathsf{Fac}_0\,a:=w_0$
$j:=0$
\vspace{1mm}
\While{\label{While2}$w_j$ is not a simple cycle nor a simple path}{
Traverse $w_j$ from start to finish\\
\vspace{.5mm}
\eIf{$w_j$ is open}{Start the traversal on $h(w_j)$}{Start the traversal on the first internal vertex of $w_j$}
Upon arriving at the earliest vertex $\eta$ that $w_j$ visits at least twice, define :\\
\vspace{.5mm}
$s_{j+1}:=(\eta_{\textrm{first}}\cdots\eta_{\textrm{last}})$
\hspace{5mm}\CommentSty{\% Cycle from the first to the last occurrence of $\eta$ in $w_j$}\\
\vspace{.5mm}
$w_{j+1}:=w_j\,/\,\{s_{j+1}\rightarrow(\eta)\}$
\hspace{5mm}\CommentSty{\% Replace $s_{j+1}$ with $(\eta)$ in $w_j$}\\
\vspace{.5mm}
$\mathsf{Fac}_{j+1}\,a:=\mathsf{Fac}_{j}\,a\,/\,\{w_j\rightarrow(w_{j+1}\odot s_{j+1})\}$
\hspace{5mm}\CommentSty{\% Replace $w_{j}$ with $w_{j+1}\odot s_{j+1}$ in $\mathsf{Fac}_j\,a$}\\
\vspace{.5mm}
$j= j+1$
}
Let $m:=j$ and $r:=w_{m}$. \label{Observation}Observe that $w_m$ is irreducible and that $\mathsf{Fac}_m\,a=\Big(~\big((r\odot s_{m})\odot s_{m-1}\big)\odot\cdots\Big)\odot s_1$.\\
$\mathsf{Fac}\,w\rightarrow \mathsf{Fac}\,w\,/\,\{a\rightarrow \mathsf{Fac}_m\,a\}$
\hspace{5mm}\CommentSty{\% Replace $a$ with $\mathsf{Fac}_m\,a$ in $\mathsf{Fac}\,w$}
}
}
\vspace{.5mm}Prime factorisation of $w$ $\equiv \mathsf{Fac}\,w$
\hspace{5mm}\CommentSty{\% All the factors appearing in $\mathsf{Fac}\,w$ are  prime}
\caption{\mystrut{16pt}Prime factorisation of individual walks}\label{AlgoCanFac}
\end{algorithm}

\begin{proof}
\sjt{We first verify the correctness of the factorisations that Algorithm \ref{AlgoCanFac} performs, and second show that for any finite-length walk, the algorithm terminates and yields the prime factorisation.}

\sjt{\textbf{Initialization:} let $\mathsf{Fac}\,w$ be any factorisation of a walk $w$ on $\G$ of finite length $\ell(w)$ and let $a\in S_{\!R}\big(\mathsf{Fac}\,w\big)$.}

\textbf{If} $a$ is a cycle off $\mu$, and $\mu$ appears $k>0$ times \sjt{as an internal} vertex of $a$, then the algorithm splits the vertex string of $a$ at each internal appearance of $\mu$, thus producing $k-1$ \sjt{cycles $c_1, c_2,\ldots, c_{k-1}$ off $\mu$. Then by construction, $a=c_{1}\conc\cdots\conc c_{k-1}$, where $\conc$ is the concatenation operator.} Since nesting coincides with concatenation over the cycles, we have $c_{j}\conc\,c_{j+1}=c_{j}\odot c_{j+1}$ and
\begin{equation}\label{cycleSPLIT}
a=c_{1}\odot\cdots\odot c_{k-1},
\end{equation}
as claimed in the algorithm. Note that \sjt{each of the cycles $c_j$ is strictly shorter than $a$ and, by construction, does not have $\mu$ as an internal vertex.}

\textbf{Else}, let $\alpha_0\alpha_1\cdots \alpha_\ell$ be the vertex sequence of $a$. \sjt{Let $\eta$ be the earliest vertex that $a$ visits at least twice (or the earliest internal vertex, if $a$ is a cycle). Let $i$ and $f$ be the indices of the earliest and latest occurrences of $\eta$ in $a$, so that $\alpha_i=\alpha_f=\eta$, and set $s_{1}$ be the string of vertices $\eta_i\eta_{i+1}\cdots\eta_f$. Let $w_1$ be the walk obtained from $a$ by replacing $s_1$ by $\eta$. Then the couple $(w_1,\,s_{1})$ is nestable, since all vertices \sjt{$\alpha_{k}$ (where $k < i$)} are visited precisely once by $w$ and therefore cannot be visited by $s_{1}$. Then, by construction of $s_{1}$ and $w_1$, we have $\mathsf{Fac}_1\,a=w_1\odot s_{1}$.
By applying the same \sjt{reasoning} for the earliest vertex $\lambda$ visited at least twice by $w_1$ (the earliest internal vertex, if $w_1$ is a cycle), we construct a nestable couple $(w_2,\,s_2)$ with $w_1=w_{2}\odot s_{2}$ and thus $\mathsf{Fac}_2\,a=(w_{2}\odot s_{2})\odot s_{1}$. Proceeding similarly with $w_2$ and all the subsequent non-irreducible walks $w_j$ thus yields}
\begin{equation}\label{eq:FactorW}
\mathsf{Fac}_m\,a=\Big(~\big((r\odot s_{m})\odot s_{m-1}\big)\odot\cdots\Big)\odot s_1,
\end{equation}
where and $r\equiv w_m$ must be irreducible, as claimed \plg{on Line \ref{Observation} of the algorithm.}

\sjt{In either of the two cases above, $a$ is} factorised into nesting products of strictly shorter \sjt{walks which are either irreducible, or will in turn be factorised into nesting products of strictly shorter walks in a later step. After at most $\ell(w) - 1$ recursive factorizations, where $\ell(w)$ is the length of the original walk $w$,} all factors obtained are either irreducible or of length $1$. Since all walks of length 1 are irreducible, it follows that the algorithm factors \sjt{any walk of finite length into a nesting product of irreducible walks} in a finite number of steps. Finally, by Lemmas \ref{irreducible} and \ref{PrimeProof}, irreducible walks are \sjt{primes. The algorithm thus yields the prime factorisation of $w$.}\qed
\end{proof}

\begin{example}[The prime factorisation of a walk]\label{examplefac} In this example we give a detailed step-by-step example illustrating \sjt{the application of Algorithm} \ref{AlgoCanFac}.
Let $\G$ be the complete ordinary (i.e. undirected) graph on $4$ vertices with vertex labels \{1,2,3,4\}, and \sjt{consider factorising the walk $w=133112343442333$.}

\vspace{1mm}Initially, the walk factorisation is simply $\mathsf{Fac}\,w:=w$ and its set of \sjt{reducible factors} is therefore $S_{\!R}\big(\mathsf{Fac}\,w\big)=w$. Since $w$ is the only factor in $\mathsf{Fac}\,w$, we let $a=w$.
\sjt{Then when Algorithm \ref{AlgoCanFac} is run, the while loop beginning on Line \ref{While1} is executed a total of 5 times:}
\begin{itemize}
\item[(1)]\sjt{Since $a$ is open, we take $w_0=a$, so that $\mathsf{Fac}_0\,a=w_0$.} Traversing $w_0$ from \sjt{left to right, we find that vertex 1 is the earliest vertex visited at least twice by $w_0$. Then $s_1=13311$, $w_1=12343442333$ and $\mathsf{Fac}_1\,a=w_1\odot s_1 = 12343442333 \odot 13311$. We now enter the while loop of Line \ref{While2}.}
\begin{itemize}
\item[(1a)]\sjt{$w_1$ is neither a simple path nor a simple cycle, and is open.} Traversing $w_1$ from its first vertex onwards \sjt{shows that vertex 2 is the} earliest vertex visited at least twice by $w_1$. Then $s_2=2343442$, $w_2=12333$ and $\mathsf{Fac}_2\,a=(w_2\odot s_2)\odot s_1$.
\item[(1b)]\sjt{$w_2$ is neither a simple path nor a simple cycle, and is open. Further, vertex 3 is the earliest vertex visited at least twice by $w_2$.} Then $s_3=333$, $w_3=123$ and $\mathsf{Fac}_3\,a=\big((w_3\odot s_3)\odot s_2\big)\odot s_1$.
\item[(1c)] \sjt{Since $w_3$ is a simple path, we exit the while loop of Line \ref{While2} and update $\mathsf{Fac}\,w$ by replacing $a$ by $\mathsf{Fac}_3\,a$. We thus obtain}
\begin{equation}
\mathsf{Fac}\,w=\Big((w_3\odot s_3)\odot s_2\Big)\odot s_1=\Big((123\odot333)\odot2343442\Big)\odot 13311.\nonumber
\vspace{.5mm}
\end{equation}
\end{itemize}
\item[(2)] \sjt{The set of reducible factors in $\mathsf{Fac}\,w$ is $S_{\!R}\big(\mathsf{Fac}\,w\big)=\{333,\,2343442\,,13311\}$. We return to the beginning of the first while loop on Line \ref{While1}, and choose $a=333$.} This walk is a cycle off 3 \sjt{and visits vertex 3 a total of $k = 3$ times. We define $c_1=33$ and $c_2=33$, and update $\mathsf{Fac}\,w$ by replacing $333$ by $33\odot 33=33^2$. The factorisation of $w$ becomes}
\begin{equation}
\mathsf{Fac}\,w=\Big((123\odot33^2)\odot2343442\Big)\odot 13311.\nonumber
\vspace{.5mm}
\end{equation}
\item[(3)] $S_{\!R}\big(\mathsf{Fac}\,w\big)=\{2343442\,,13311\}$. We choose $a=13311$. \sjt{Then $a$ is a cycle off 1 and visits vertex 1 a total of $k = 3$ times. We define $c_1=1331$ and $c_2=11$, and update $\mathsf{Fac}\,w$ by replacing $a$ with $c_1\odot c_2$, yielding}
\begin{equation}
\mathsf{Fac}\,w=\Big((123\odot33^2)\odot2343442\Big)\odot (1331\odot11).\nonumber
\end{equation}
\item[(4)] $S_{\!R}\big(\mathsf{Fac}\,w\big)=\{2343442\,,1331\}$. \sjt{We choose $a=1331$. Now $a$ is a cycle off 1 that does not have 1 as an internal vertex. Executing the second while loop on Line \ref{While2} results in $s_1=33$, $w_1=131$ and $\mathsf{Fac}_1\,a=w_1\odot s_1$.}
\begin{itemize}
\item[(4a)] \sjt{Since $w_1$ is a simple cycle, we exit the while loop of Line \ref{While2}} and update $\mathsf{Fac}\,w$, replacing $a$ with $\mathsf{Fac}_1\,a$. We obtain
\begin{equation}
\mathsf{Fac}\,w=\Big((123\odot33^2)\odot2343442\Big)\odot \Big((131\odot33)\odot 11\Big).\nonumber
\end{equation}
\vspace{.5mm}
\end{itemize}
\item[(5)] \sjt{The set of reducible factors has been reduced to $S_{\!R}\big(\mathsf{Fac}\,w\big)=\{2343442\}$. We therefore set $a=2343442$, which is a cycle off 2 that does not have 2 as an internal vertex.} Then $s_1=343$, $w_1=23442$ and $\mathsf{Fac}_1\,a=w_1\odot s_1$.
\begin{itemize}
\item[(5a)] \sjt{$w_1$ is neither a simple path nor a simple cycle, but a cycle off 2 that does not have 2 as internal vertex. Then the first pass through the while loop of Line \ref{While2} yields $s_2=44$, $w_2=2342$ and $\mathsf{Fac}_2\,a=(w_2\odot s_2)\odot s_1$.}
\item[(5b)] \sjt{$w_2$ is a simple cycle. We exit the while loop of Line \ref{While2} and update $\mathsf{Fac}\,w$, replacing $a$ with $\mathsf{Fac}_2\,a$.} We obtain
\begin{equation}\label{eq:walkFactored}
\mathsf{Fac}\,w=\bigg((123\odot 33^2)\odot \Big((2342\odot 44)\odot 343\Big)\bigg)\odot \Big((131\odot33)\odot 11\Big).
\vspace{-1mm}
\end{equation}
\end{itemize}
\end{itemize}
At this point $S_{\!R}\big(\mathsf{Fac}\,w\big)$ is empty, \plg{we exit the while loop of Line \ref{While1},} and Eq.~(\ref{eq:walkFactored}) is the prime factorisation of $w$ into nesting products of prime walks.
A pictorial representation of the operations performed by the algorithm is given in Fig.~\ref{fig:algo}.
\begin{figure}[t!]
\hfill
\subfloat{\begin{minipage}{0.41\textwidth}
\includegraphics[width=1\textwidth]{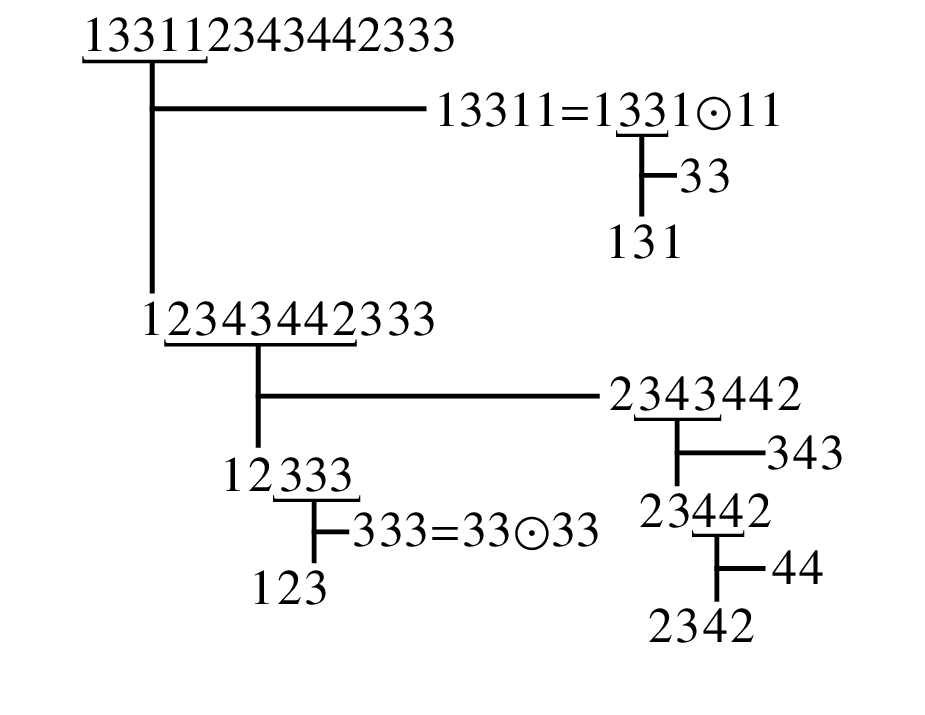}
\end{minipage}\label{fig:procedure}}
\hspace{-5mm}
\begin{minipage}{0.35\textwidth}
\subfloat{\includegraphics[width=1.2\textwidth]{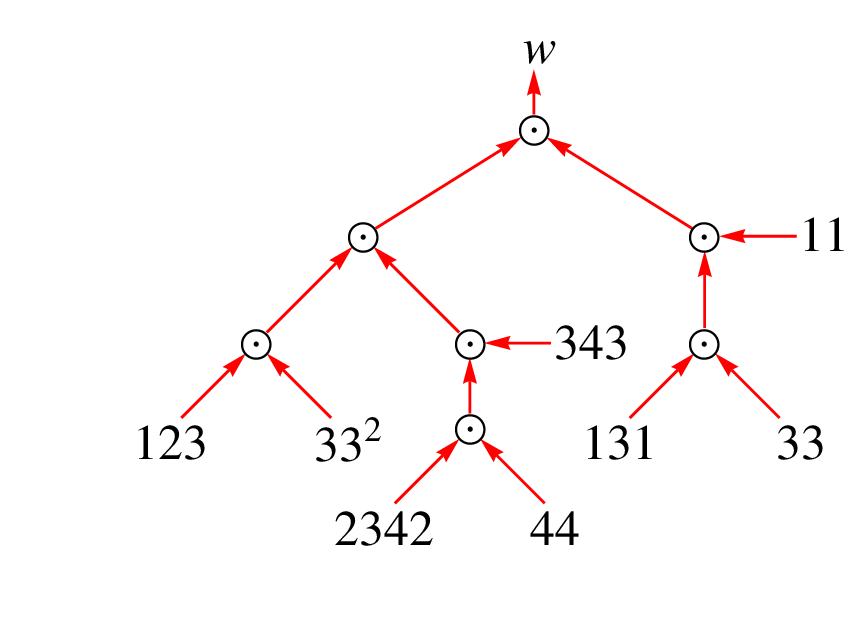}}
\end{minipage}
\hspace{20mm}
\caption{\sjt{(left) a schematic representation of the steps of Algorithm \ref{AlgoCanFac} as outlined in Example \ref{examplefac}; (right) a tree $T_w$ representing} the prime factorisation of $w$. Each node corresponds to a nesting product, the leaves are the irreducible factors of $w$, and the root is the walk $w$ itself. The tree $T_w$ is in fact a subgraph of the Hasse diagram of the set of walks partially ordered by the divisibility relation of Definition \ref{def:divisibility}. This observation lies at the heart of a \sjt{``number theory" of prime walks, which will be presented in detail elsewhere.}}\label{fig:algo}
\end{figure}
\end{example}

\subsection{Prime factorisation of walk sets}\label{SectionFactorBasis}
Concluding this section on unique factorisation, we establish the factorisation of sets of walks into nested sets of primes. More precisely, we obtain \sjt{an expression for the set of all walks between any two vertices of $\G$ in terms of Kleene stars} of nested sets of simple paths and simple cycles. This result will prove decisive in the next section, when we obtain the prime representation of any series of walks.

\begin{remark}[Nesting sets]
Let $A$ and $B$ be two sets of walks on $\G$. Then we write $A\odot B$ for the set obtained by nesting every element of $B$ into every element of $A$.
\end{remark}

\begin{remark}[Kleene star and nesting Kleene star]
Let $\alpha$ be a vertex \sjt{on $\G$ and} $E_\alpha\subseteq W_{\G;\,\alpha\alpha}$ be a subset of the set of all walks from $\alpha$ to itself on $\G$. Set $E_{\alpha}^0 = \{(\alpha)\}$ and $E_{\alpha}^{i} = E_{\alpha}^{i-1}\conc E_{\alpha}$ for $i\ge 1$. Then the \emph{Kleene star} of $E_{\alpha}$, denoted $E_{\alpha}^\ast$, is the set of walks formed by concatenating any number of elements \sjt{of $E_{\alpha}$: that is,} $E_{\alpha}^{\ast}= \bigcup_{i=0}^\infty E_{\alpha}^i$ \cite{Ebbinghaus1994}. The \emph{nesting Kleene star} of $E_{\alpha}$, denoted $E_{\alpha}^{\odot\ast}$, is the equivalent of the Kleene star \sjt{with concatenation replaced} by the nesting product: $E_{\alpha}^{\odot\ast}=\bigcup_{i=0}^\infty E_{\alpha}^{\odot i}$ where $E_{\alpha}^{\odot 0}=\{(\alpha)\}$ and $E_{\alpha}^{\odot i} = E_{\alpha}^{\odot(i-1)}\odot E_{\alpha}$ for $i\geq 1$. Since nesting coincides with concatenation for cycles off the same vertex, the nesting \sjt{Kleene star coincides} with the \sjt{usual Kleene star, that is} $E_{\alpha}^{\odot\ast}=E_{\alpha}^{\ast}$. \sjt{From now on we therefore do} not distinguish between the two.
\end{remark}

\begin{theorem}[Factorisation of walk sets]\label{FactorBasis}
Let $\nu_0$ and $\nu_p$ be two vertices on $\G$. Then the set of all walks on $\G$ from $\nu_0$ to $\nu_p$ is expressible solely in terms of sets of prime walks on $\G$. This expression is \sjt{given} by the following recursive relations:
\begin{subequations}\label{eqn:FactorBasis}
\begin{align}
&W_{\G;\,\nu_0\nu_p}=\bigg(\Big(~(\,\Pi_{\G;\,\nu_0\nu_p}\odot   C^\ast_{\G\backslash\{\nu_0,\cdots,\nu_{\ell(p)-1}\};\,\nu_p})\odot\cdots\odot   C^\ast_{\G\backslash\{\nu_0\};\,\nu_1}\Big)\odot  C^\ast_{\G;\,\nu_0}\bigg)\,\label{WG},\\
\shortintertext{where $\nu_0\nu_1\cdots\nu_{p-1}\nu_p\in\Pi_{\G;\,\nu_0\nu_p}$ is a simple path and}
&C_{\G;\,\mu_c}=\bigg(\Big(\,(\,\Gamma_{\G;\,\mu_c}\odot   C^\ast_{\G\backslash\{\mu_c,\,\mu_1,\cdots,\,\mu_{c-2}\};\,\mu_{c-1}})\odot\cdots\odot C^\ast_{\G\backslash\{\mu_c,\,\mu_1\};\,\mu_2}\Big)\odot  C^\ast_{\G\backslash\{\mu_c\};\,\mu_1}\bigg),\label{AG}
\end{align}
\end{subequations}
with $\mu_c\mu_1\cdots\mu_{c-1}\mu_c\in\Gamma_{\G;\,\mu_c}$ is a simple cycle.
\end{theorem}

\noindent Note that if $\nu_0=\nu_p$, \sjt{then $\Pi_{\G;\,\nu_0\nu_0}=\big\{(\nu_0)\big\}$} and $W_{\G;\,\nu_0\nu_0}=C_{\G;\,\nu_0}^\ast$ with $C_{\G;\,\nu_0}$ given by Eq.~(\ref{AG}). This is the factorization of sets of cycles on $\G$.

\vspace{2mm}\plg{We show in the proof of the Theorem that the set $C_{\G;\,\mu_c}$ is the set of cycles from $\mu_c$ to itself on $\G$ that do not have $\mu_c$ as an internal vertex. This set is factorized} recursively through Eq.~(\ref{AG}). Indeed $C_{\G;\,\mu_c}$ is expressed in terms of $C_{\G\backslash\{\mu_c,\,\mu_1,\cdots,\,\mu_{j-1}\};\,\mu_{j}}$ which is in turn factorized through Eq.~(\ref{AG}) but on the subgraph $\G\backslash{\{\mu_c,\ldots,\mu_{j-1}}\}$ of $\G$. The recursion stops when vertex $\mu_j$ has no neighbour on this subgraph, in which case $C_{\G\backslash\{\mu_c,\,\mu_1,\cdots,\,\mu_{j-1}\};\,\mu_{j}}=\Gamma_{\G\backslash\{\mu_c,\,\mu_1,\cdots,\,\mu_{j-1}\};\,\mu_{j}}=\{(\mu_j\mu_j)\}$ if the loop $(\mu_j\mu_j)$ exists and \sjt{$C_{\G\backslash\{\mu_c,\,\mu_1,\cdots,\,\mu_{j-1}\};\,\mu_{j}}=\big\{(\mu_j)\big\}$ otherwise.}
The maximum depth at which this recursion stops is discussed in \S \ref{StarHeightsection}.

\begin{proof}
Let $\nu_0$ and $\nu_p$ be two connected vertices of $\G$, and consider $w$ a walk from $\nu_0$ to $\nu_p$.
For convenience, we define
\begin{align}
B_{\G;\,\nu_0\nu_p}&:=\bigg(\Big(~(\,\Pi_{\G;\,\nu_0\nu_p}\odot   A^\ast_{\G\backslash\{\nu_0,\cdots,\nu_{p-1}\};\,\nu_p})\,\odot\cdots\odot   A^\ast_{\G\backslash\{\nu_0\};\,\nu_1}\Big)\odot  A^\ast_{\G;\,\nu_0}\bigg),
\end{align}
where $A_{\G\backslash\{\nu_0,\cdots,\nu_{j-1}\};\,\nu_j}$ designates the set of cycles off $\nu_j$ on $\G\backslash\{\nu_0,\cdots,\nu_{j-1}\}$ that do not have $\nu_j$ \sjt{as an internal vertex.}
First, we will show that $B_{\G;\,\nu_0\nu_p}=W_{\G;\,\nu_0\nu_p}$ by showing that $W_{\G;\,\nu_0\nu_p}\subseteq B_{\G;\,\nu_0\nu_p}$ and $B_{\G;\,\nu_0\nu_p}\subseteq W_{\G;\,\nu_0\nu_p}$. In a second time, we will show that $A_{\G;\,\mu_c}$ identifies with the set $C_{\G;\,\mu_c}$ in \sjt{Theorem \ref{FactorBasis}.}

\vspace{.5mm}By Eq.~(\ref{eq:FactorW}), $w$ can be expressed as a simple path $r\in \Pi_{\G;\,\nu_0\nu_p}$ with \sjt{a collection of cycles $s_{j}$ nested into it:} that is
$\mathsf{Fac}\,w:=\big(~((r\odot s_{m})\odot s_{m-1})\odot\cdots\big)\odot s_1$ is a valid factorisation of $w$.
By construction, the $s_j$ \sjt{for $1 \le j \le m$ are cycles} nested off different vertices of the simple path $r$. \sjt{For each vertex} $\nu_k$ of $r$, we define $s_{\nu_k}=(\nu_k)$ if no $s_j$ is nested off $\nu_k$ and $s_{\nu_k}=s_j$ if $h(s_j)=\nu_k$. Then let
\begin{equation}\label{factoriDEMO}
\mathsf{Fac}'(w):=\Big(~\big((r\odot s_{\nu_p})\odot s_{\nu_{p-1}}\big)\odot\cdots\Big)\odot s_{\nu_0},
\end{equation}
and note that $\mathsf{Fac}'(w)\equiv \mathsf{Fac}\,w$.
By the nestable property, $s_{\nu_j}$ cannot visit \sjt{any of $\nu_0,\cdots,\nu_{j-1}$} and must therefore \sjt{be an element of $W_{\G\backslash\{\nu_0,\cdots,\nu_{j-1}\};\,\nu_j\nu_j}$.} By Eq.~(\ref{cycleSPLIT}), \sjt{any element of} $W_{\G\backslash\{\nu_0,\cdots,\nu_{j-1}\};\,\nu_j\nu_j}$ can be decomposed into nesting products of shorter cycles $c_i$ off $\nu_j$ that do not have $\nu_j$ \sjt{as an internal vertex.} Therefore  $s_{\nu_j}\in A_{\G\backslash\{\nu_0,\cdots,\nu_{j-1}\};\,\nu_j}^\ast$ and \sjt{consequently $w\in B_{\G;\,\nu_0\nu_p}$. Since $w$ was arbitrary, it follows that $W_{\G;\,\nu_0\nu_p}\subseteq B_{\G;\,\nu_0\nu_p}$.} Furthermore, any element of $B_{\G;\,\nu_0\nu_p}$ is a walk on $\G$ from $\nu_0$ to $\nu_p$\sjt{, so that $B_{\G;\,\nu_0\nu_p}\subseteq W_{\G;\,\nu_0\nu_p}$.} Hence we deduce $B_{\G;\,\nu_0\nu_p}= W_{\G;\,\nu_0\nu_p}$.

\vspace{.5mm}It remains to show that the set $C_{\G;\,\mu_c}$ of \sjt{Eq.~\eqref{eqn:FactorBasis}} is $A_{\G;\,\mu_c}$. Let $c\in A_{\G;\,\mu_c}$. Applying the same reasoning as above, $c$ factorises as in Eq.~(\ref{factoriDEMO}), but with \sjt{$r$ being a simple cycle (i.e.~an element of $\Gamma_{\G;\,\mu_c}$) instead of a simple path. Thus $c$ is an element of the set}
\begin{equation}\label{ensemright}
\bigg(\Big(~(\,\Gamma_{\G;\,\mu_c}\odot   A^\ast_{\G\backslash\{\mu_c,\,\mu_1,\cdots,\,\mu_{c-2}\};\,\mu_{c-1}})\,\odot\cdots\odot A^\ast_{\G\backslash\{\mu_c,\,\mu_1\};\,\mu_2}\Big)\odot  A^\ast_{\G\backslash\{\mu_c\};\,\mu_1}\bigg).
\end{equation}
\sjt{Any element of this set is a cycle off $\mu_c$ that does not have $\mu_c$ as an internal vertex. Consequently, $A_{\G;\,\mu_c}$ identifies with the set of Eq.~(\ref{ensemright}). If $\mu_c$ has no neighbour on $\G$, then $A_{\G;\,\mu_c}=\big\{(\mu_c\mu_c)\big\}$ if the loop $(\mu_c\mu_c)$ exists and $A_{\G;\,\mu_c}=\big\{(\mu_c)\big\}$} otherwise. \sjt{Thus $A_{\G;\,\mu_c}$ and $C_{\G;\,\mu_c}$ both fulfill the same recursive relation and value on vertices with no neighbour and it follows that they are equal.}  This establishes Eq.~(\ref{AG}) and, together with $B_{\G;\,\nu_0\nu_p}= W_{\G;\,\nu_0\nu_p}$, Eq.~(\ref{WG}). \qed
\end{proof}

\vspace{2mm}\begin{example}[Prime factorisation of a walk set]\label{exampleK3}
\sjt{Let $\mc{T}_{3}$ be the complete graph on three vertices, with a self-loop on each vertex. We label the vertices 1, 2, and 3. In this example we derive the prime factorisation of the set of all walks from 1 to 1 on $\mc{T}_3$.}
\sjt{The set of all walks from 1 to 1 is a set of cycles, and Eq.~(\ref{WG}) thus yields} $W_{\mc{T}_{3};\,11}=C_{\mc{T}_{3};\,11}^\ast$. To factorise $C_{\mc{T}_{3};\,11}^\ast$, we note that the set of simple cycles from 1 to itself \sjt{is $\Gamma_{\mc{T}_{3};\,1}=\{11,121,131,1231,1321\}$.} Thus Eq.~(\ref{AG}) gives
\begin{align}\label{FactorW11K3step1}
W_{\mc{T}_{3};\,11}&=\bigg\{11,\,121\odot C^\ast_{\mc{T}_{3}\backslash\{1\};\,22},\,131\odot C^\ast_{\mc{T}_{3}\backslash\{1\};\,33},\\
&\hspace{0mm}\big(1231\odot C^\ast_{\mc{T}_{3}\backslash\{1,2\};\,33}\big)\odot C^\ast_{\mc{T}_{3}\backslash\{1\};\,22},\,\big(1321\odot C^\ast_{\mc{T}_{3}\backslash\{1,3\};\,22}\big)\odot C^\ast_{\mc{T}_{3}\backslash\{1\};\,33}\bigg\}^\ast.\nonumber
\end{align}
We now \sjt{use Eq.~(\ref{AG}) to factor each of the sets $C^\ast_{\G;\mu\mu}$.} Since \sjt{$\Gamma_{\mc{T}_{3}\backslash\{1\};\,2}=\{22,232\}$ and $\Gamma_{\mc{T}_{3}\backslash\{1,3\};\,2}=\{22\}$,} we have
\begin{equation}\label{FactorW11K3step2}
C^\ast_{\mc{T}_{3}\backslash\{1\};\,22}=\{22,232\odot C^\ast_{\mc{T}_{3}\backslash\{1,2\};\,33}\}^\ast\quad\text{and}\quad C^\ast_{\mc{T}_{3}\backslash\{1,3\};\,22}=\{22\}^\ast,
\end{equation}
and the analogous expressions produced by exchanging the labels $2$ and $3$. Inserting these
expressions into Eq.~(\ref{FactorW11K3step1}), we arrive at
\begin{align}\label{W11K3Factor}
W_{\mc{T}_{3};\,11}=\bigg\{&11,121\odot\Big\{22,232\odot\{33\}^\ast\Big\}^\ast,\,131\odot\Big\{33,323\odot\{22\}^\ast\Big\}^\ast,\\
&\hspace{-5mm}\big(1231\odot\{33\}^\ast\big)\odot\Big\{22,232\odot\{33\}^\ast\Big\}^\ast,\,\big(1321\odot\{22\}^\ast\big)\odot\Big\{33,323\odot\{22\}^\ast\Big\}^\ast\bigg\}^\ast.\nonumber
\end{align}
This set contains the prime factorisation of any cycle off $1$ on $\mc{T}_3$.
\end{example}

\section{Prime factorisation of series of walks}\label{PathSumPrimeSeries}
The main interest of the existence and uniqueness of the prime factorisation of walks \sjt{is that it permits a series of walks on a digraph $\G$ to be resummed into an expression that involves only the prime elements of $\G$.} This is analogous to how the fundamental theorem of arithmetic \sjt{leads to the existence} of Euler products for the Riemann zeta function and other totally multiplicative \sjt{functions on the integers}. In fact, \sjt{as we will show in a future work,} the relation between these two cases is not simply an analogy but can be established rigorously.

In this section, we begin by obtaining an explicit closed-form expression involving only prime walks for the formal series of all walks on the graph. \sjt{We also obtain the equivalent expression for series of walk weights on a weighted directed graph. These resummed walk series have found applications} in linear algebra \cite{Giscard2011b}, machine learning \cite{Malioutov2006} and physics \cite{Brydges1983}, \cite{Giscard2014}.

\subsection{Formal series of walks}\label{formal}
Let $\alpha$ and $\omega$ be two vertices on $\G$. The \emph{characteristic series} of the set $W_{\G;\,\alpha\omega}$ of all walks from $\alpha$ to $\omega$ on $\G$ is the formal series \cite{Berstel2008}
\begin{equation}
\Sigma_{\G;\,\alpha\omega}:=\sum_{w\in W_{\G;\,\alpha\omega}} w.
\end{equation}
In other words, the coefficient of $w$ in $\Sigma_{\G;\,\alpha\omega}$, denoted $(\Sigma_{\G;\,\alpha\omega},w)$, is 1 if $w\in W_{\G;\,\alpha\omega}$ and 0 otherwise.


By using the fact that every open walk can be factorised into a simple path and a collection of nested cycles, we rewrite $\Sigma_{\G;\,\alpha\omega}$ as a series over simple paths by modifying each path in the series to include all collections of cycles that can be nested off the vertices it visits. To preserve the vertex-edge notation of walks, we implement this modification by replacing each vertex $\alpha$ in a simple path by a `dressed vertex' $(\alpha)'_{\G}$ defined to represent the characteristic series of all cycles that can be nested off $\alpha$ on $\G$:
\begin{equation}\label{DressedVertex}
(\alpha)'_{\G}:=\sum_{c\,\in\, W_{\G;\,\alpha\alpha}}c=\Sigma_{\G;\,\alpha\alpha}.
\end{equation}
\sjt{We rewrite this characteristic series as a series over simple cycles $\gamma\in\Gamma_{\G;\,\alpha}$ by replacing each vertex $\mu$ visited} by a simple cycle $\gamma$ by a dressed vertex representing the characteristic series of all the cycles that can be nested off $\mu$ on the appropriate subgraph of $\G$. \sjt{Applying this approach recursively yields} a representation of the formal series $\Sigma_{\G;\,\alpha\omega}$ which only involves simple paths and simple cycles \sjt{(i.e.~the prime walks on $\G$).}


\begin{theorem}[Formal path-sum]\label{FactorSum}
Using the vertex-edge notation for walks, the formal characteristic series of all walks from $\alpha$ to $\omega$ on $\G$ has the following \sjt{expression, which involves only primes on $\G$:}
\begin{subequations}
\begin{align}\label{eqn:SumOfAllWalks}
\Sigma_{\G;\,\alpha\omega}=\sum_{\Pi_{\G;\,\alpha\omega}}\left(\alpha\right)'_\G\left(\alpha\nu_1\right) \left(\nu_1\right)'_{\G\backslash\{\alpha\}} \cdots (\nu_{\ell(p)-1}\omega)\left(\omega\right)'_{\G\backslash\{\alpha,\nu_1,\ldots,\nu_{\ell(p)-1}\}},
\end{align}
where $p=(\alpha\nu_1\cdots\nu_{\ell(p)-1}\omega)$ is a simple path of length $\ell(p)$ from $\alpha\equiv \nu_0$ to $\omega\equiv \nu_{\ell(p)}$, and $\left(\alpha\right)'_\G$ denotes the dressed vertex $\alpha$ on $\G$, defined as the formal series of all cycles off $\alpha$ on $\G$ and given explicitly by
\begin{equation}
\left(\alpha\right)'_\G
=\Bigg[\left(\alpha\right) -\!\!\!\sum_{\gamma\in\Gamma_{\G;\,\alpha}}
 \left(\alpha\right)\left(\alpha\mu_1\right)\left(\mu_1\right)'_{\G\backslash\{\alpha\}}\left(\mu_1\mu_2\right)\cdots(\mu_{\ell(\gamma)-1})'_{\G\backslash\{\alpha,\mu_1,\ldots,\mu_{\ell(\gamma)-1}\}} (\mu_{\ell(\gamma)-1}\alpha)(\alpha)\Bigg]^{-1}\!\!,\label{eqn:DressedVertexClosedForm2B}
\end{equation}
\end{subequations}
with $\gamma=(\alpha\mu_1\cdots\mu_{\ell(\gamma)-1}\alpha)$ a simple cycle of length $\ell(\gamma)$ off $\alpha$.
\end{theorem}
\sjt{The formal series $\Sigma_{\G;\alpha\omega}$ is expressed recursively in terms of formal series on subgraphs of $\G$. We term these formal series the dressed vertices, and denote them by e.g.~ $(\mu_{j})'_{\G\backslash\{\alpha,\,\mu_1,\,\cdots,\,\mu_{j-1}\}}$. These subseries are in turn obtained through Eq.~(\ref{eqn:DressedVertexClosedForm2B}), but on the subgraphs of $\G$ (e.g. $\G\backslash\{\alpha,\mu_1,\ldots,\mu_{j-1}\}$ in the case of $(\mu_{j})'_{\G\backslash\{\alpha,\,\mu_1,\,\cdots,\,\mu_{j-1}\}}$).} The recursion stops when vertex $\mu_j$ has no neighbour on this subgraph. \sjt{In this case the dressed vertex is given by
\begin{align}
  (\mu_{j})'_{\G\backslash\{\alpha,\,\mu_1,\,\cdots,\,\mu_{j-1}\}}=\sum_{n\geq0}(\mu_j\mu_j)^n= \begin{cases}
    [(\mu_j)-(\mu_j\mu_j)]^{-1}&\text{if the loop $(\mu_j\mu_j)$ exists},\\
    (\mu_j) & \text{otherwise,}
  \end{cases}
\end{align}
where $(\mu_j)$ is the trivial walk off the vertex $\mu_j$.}\\

\vspace{1mm}The recursive nature of Eq.~(\ref{eqn:DressedVertexClosedForm2B}) implies that the result of Theorem \ref{FactorSum} for $\Sigma_{\G;\,\alpha\omega}$ yields a formal continued fraction involving only prime walks. On finite digraphs, the depth of this continued fraction is finite but determining its precise value is difficult, \sjt{as we discuss further in \S\ref{StarHeightsection}.}

\begin{proof}
\sjt{Theorem \ref{FactorSum}} follows from the factorisation of sets of walks into nested sets of primes presented in Theorem \ref{FactorBasis}. We provide two proofs of the theorem, one based on  formal series, the other on quivers.

\textit{Proof 1} \sjt{Consider the set of all walks from $\alpha$ to $\omega$ on $\G$, denoted by $W_{\G;\,\alpha\omega}$.} We first decompose $W_{\G;\,\alpha\omega}$ using Eq.~(\ref{WG}), identifying $\alpha$ with $\nu_0$ and $\omega$ with $\nu_{\ell(p)}$ for convenience, then sum over the elements of the sets on both sides of the equality. This yields, in vertex-edge notation,
\begin{align}\label{SumStep0}
\Sigma_{\G;\,\alpha\omega}&=\sum_{p\in \Pi_{\G;\,\alpha\omega}}\left(\sum_{c_0\in C^\ast_{\G;\,\alpha}}c_0\right)(\alpha\nu_1)\left(\sum_{c_1\in C^\ast_{\G\backslash\{\alpha\};\,\nu_1}}c_1\right)(\nu_1\nu_2)\cdots\\
&\hspace{50mm}\cdots(\nu_{\ell(p)-1}\omega)\left(\sum_{c_{\ell(p)}\in C^\ast_{\G\backslash\{\alpha,\nu_1,\cdots,\nu_{\ell(p)-1}\};\,\omega}}c_{\ell(p)}\right),\nonumber
\end{align}
which we obtain upon nesting the sets $C^\ast_{\G;\,\alpha},\,C^\ast_{\G\backslash\{\alpha\};\,\nu_1},\ldots$ into the \sjt{simple path $p = \alpha\nu_1\cdots\nu_{\ell(p)-1}\omega \in \Pi_{\G;\,\alpha\omega}$} at the appropriate positions. Equation (\ref{SumStep0}) shows that \sjt{the sum over each of these sets gives rise to an `effective vertex', which is produced by dressing a `bare vertex' $\nu_j$ by all cycles that visit it on the subgraph $\G\backslash\{\alpha,\ldots,\nu_{j-1}\}$. Motivated by this observation, we therefore define the vertex $\alpha$ dressed by cycles on $\G$, denoted by $(\alpha)'_{\G}$, to be the formal series}
\begin{equation}\label{Dress}
(\alpha)'_{\G}:=\sum_{c_0\in C^\ast_{\G;\,\alpha}}c_0.
\end{equation}
It follows that Eq.~(\ref{SumStep0}) yields Eq.~\eqref{eqn:SumOfAllWalks}, with dressed vertices representing the characteristic series \sjt{of the sets $C^\ast_{\G\backslash\{\alpha,\nu_1\cdots \nu_{j-1}\};\,\nu_j}$.} These series are proper \cite{Droste2009}: their constant term is a \sjt{trivial walk (e.g. $(\alpha)$ in Eq.~(\ref{Dress}))} which is different from 0. Thus the series represent formal inverses \cite{Sakarovitch2009}, e.g. $(\alpha)'_{\G}=[(\alpha)-\sum_{c_0\in C_{\G;\,\alpha}} c_0]^{-1}$. Note that the sum appearing in the inverse runs over $C_{\G;\,\alpha}$ rather than its Kleene star, showing that the inverse is \sjt{a representation of the characteristic series of $C^\ast_{\G;\,\alpha}$ as a geometric series in the characteristic series of $C_{\G;\,\alpha}$}.

By combining these results with Eq.~(\ref{AG}), the dressed vertices are seen to be of the form
  \begin{align}
  \left(\alpha\right)'_\G  & =\Bigg[(\alpha)-\sum_{\gamma\in\Gamma_{\G;\alpha}}(\alpha)\left(\alpha\mu_1\right)(\mu_1)'_{\G\backslash\{\alpha\}}\left(\mu_1\mu_2\right)\cdots \left(\mu_{\ell(\gamma)-1}\alpha\right)\Bigg]^{-1},\label{SumOverKleeneStars}
\end{align}
\sjt{where $\gamma=\alpha\mu_1\cdots\mu_{\ell(\gamma)-1}\alpha$ is a simple cycle} from $\alpha$ to itself and $(\alpha)$ is the left-identity common to all walks of $W_{\G;\alpha\alpha}$: for any cycle $c$ off $\alpha$, we have $c^0=(\alpha)$. In this expression, the dressed vertices again represent sums over the Kleene stars that appear when $C_{\G;\,\alpha}$ is decomposed using Eq.~(\ref{AG}). This establishes Eq.~(\ref{eqn:DressedVertexClosedForm2B}).\qed

\textit{Proof 2} We obtain the same results explicitly with the help of quivers. Let $\mathfrak{V}=\{V\}$ be a collection of vector spaces, each of arbitrary finite dimension, such that $\mathfrak{V}$ is in one to one correspondence with the vertex set $\mc{V}(\G)$ of the finite directed graph $\G$. For simplicity we designate by $V_\mu\in\mathfrak{V}$ the vector space associated to vertex $\mu\in\mc{V}(\G)$. Let $\mathfrak{F}=\{\varphi_{\nu\ot\mu}:V_\mu\to V_\nu\}$ be a collection of linear mappings in one to one correspondence with the edge set $\mc{E}(\G)$ of $\G$. We associate the linear mapping $\varphi_{\nu\ot\mu}\in\mathfrak{F}$ to the directed edge from $\mu$ to $\nu$.
Then $\mathfrak{G}=(\mathfrak{V},\,\mathfrak{F})$ is a representation of the directed graph $\G$, which in this context is also called a \emph{quiver} \cite{Dersken2005,Savage2005}. The representation of a walk \sjt{$w=\alpha_0\alpha_1\cdots\alpha_{\ell}\in W_\G$ of length $\ell$} is the linear mapping $\varphi_w$ obtained from the composition of the linear mappings representing the successive edges traversed by the \sjt{walk: that is,} $\varphi_w=\varphi_{\alpha_{\ell}\ot\alpha_{\ell-1}}\conc\cdots\conc\varphi_{\alpha_2\ot\alpha_1}\conc\varphi_{\alpha_{1}\ot\alpha_0}$. The representation of a \sjt{trivial walk $(\mu)$ is the identity map $1_{\mu}$ on $V_\mu$, and the representation of the empty walk $0$ is the 0 map.}

Now \sjt{define $\varphi_{\Gamma_{\G;\,\alpha}}$ to be the} mapping representing the finite series $\sum_{\gamma\in\Gamma_{\G;\,\alpha}}\gamma'$. By linearity, \sjt{we have} $\varphi_{\Gamma_{\G;\,\alpha}}=\sum_{\gamma\in\Gamma_{\G;\,\alpha}}\varphi_{\gamma'}$.
Define $\varphi_{(\alpha)'_{\G}}=\sum_{p\in\mathbb{N}}\varphi_{\Gamma_{\G;\,\alpha}}^{(p)}$, where $\varphi_{\Gamma_{\G;\,\alpha}}^{(p)}$ is the $p$-th composition of $\varphi_{\Gamma_{\G;\,\alpha}}$ with itself, $\varphi_{\Gamma_{\G;\,\alpha}}^{(0)}$ being the local identity map $1_{\alpha}$. Then observe that
$\varphi_{(\alpha)'_{\G}}\conc\varphi_{\Gamma_{\G;\,\alpha}}=\varphi_{\Gamma_{\G;\,\alpha}}\conc\varphi_{(\alpha)'_{\G}}=\sum_{p\in\mathbb{N}}\varphi_{\Gamma_{\G;\,\alpha}}^{(p+1)}=\varphi_{(\alpha)'_{\G}}-1_{\alpha}$. Consequently, $\varphi_{(\alpha)'_{\G}}$ is the compositional inverse
\begin{equation}
\varphi_{(\alpha)'_{\G}}=\big(1_{\alpha}-\varphi_{\Gamma_{\G;\,\alpha}}\big)^{(-1)},
\end{equation}
which is the quiver representation of the formal inverse representation of a dressed vertex $(\alpha)'_{\G}=[(\alpha)-\sum_{c\in C_{\G;\,\alpha}} c]^{-1}$.\qed
\end{proof}

\vspace{2mm}\begin{example}[Formal series of walks on a digraph]\label{FactorExample} Let $\G$ be the digraph illustrated \sjt{in Fig. \ref{fig:GRef},} and consider the formal series $\Sigma_{\G;\,11}$ of all walks from vertex 1 to itself on $\G$.
\begin{figure}[t!]
\begin{center}
\vspace{-2mm}
\includegraphics[width=.5\textwidth]{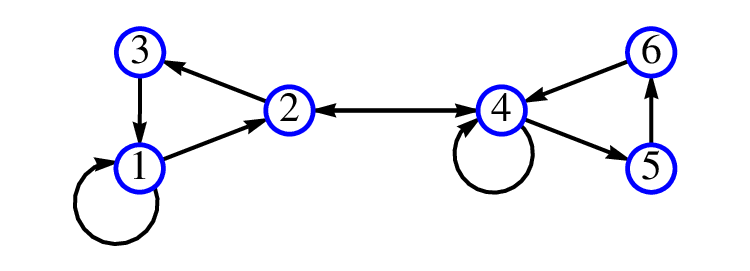}
\vspace{-3mm}
\caption[Digraph of Example \ref{FactorExample}]{The digraph $\mc{G}$ of Example \ref{FactorExample}.}
\label{fig:GRef}
\end{center}
\vspace{-2mm}
\end{figure}
Theorem \ref{FactorSum} yields this series as
\begin{equation}
\Sigma_{\G;\,11}=\Big[(1)-(11)-(12)(2)'_{\G\backslash\{1\}}(23) (3)'_{\G\backslash\{1,2\}}(31)\Big]^{-1},
\end{equation}
where we used that the set of simple cycles off 1 on $\G$ is $\Gamma_{\G;\,11}=\{11,1231\}$. Note that $\Gamma_{\G\backslash\{1,2\};33}$ is \sjt{empty, so that $(3)'_{\G\backslash\{1,2\}}=(3)$ is trivial.} We obtain the dressed vertex $(2)'_{\G\backslash\{1\}}$, which represents the sum of all cycles off $2$ on $\G\backslash\{1\}$, thanks to Eq.~(\ref{eqn:DressedVertexClosedForm2B}) as
\begin{subequations}
\begin{align}
(2)'_{\G\backslash\{1\}}&=\Big[(2)-(24)(4)'_{\G\backslash\{1,2\}}(42)\Big]^{-1},\\
\shortintertext{since $\Gamma_{\G\backslash\{1\};22}=\{(242)\}$. Similarly,}
(4)'_{\G\backslash\{1,2\}}&=\Big[(4)-(44)-(45)(5)(56)(6)(64)\Big]^{-1},
\end{align}
\end{subequations}
on using that $\Gamma_{\G\backslash\{1,2\};44}=\{(44),(4564)\}$ and the sets $\Gamma_{\G\backslash\{1,2,4\};55}$ and $\Gamma_{\G\backslash\{1,2,4,5\};66}$ are empty so that $(5)'_{\G\backslash\{1,2,4\}}=(5)$ and $(6)'_{\G\backslash\{1,2,4,5\}}=(6)$. Finally,
\begin{equation}
\Sigma_{\G;\,11}=\left[(1)-(11)-(12)\left[(2)-(24)\Big[(4)-(44)-(4564)\Big]^{-1}(42)\right]^{-1}(231)\right]^{-1}.
\end{equation}
This expression for $\Sigma_{\G;\,11}$ is recovered upon summing over the prime factorised form for the set of all walks from 1 to itself \sjt{on $\G$, namely}
\begin{equation}\label{WG11exFact}
W_{\G;\,11}=\Big\{11,\,1231\odot\big\{242\odot\{44,\,4564\}^\ast\big\}^\ast\Big\}^\ast.
\end{equation}
\end{example}

\subsection{Prime factorisations of weighted sums of walks}\label{weightedsum}
We now consider the weighted counterparts \sjt{of formal walk series,} which arise when summing walks on weighted digraphs. Evidently, \sjt{series of walk weights, rather than formal series of walks, are the objects found in applications.}

A \emph{weighted digraph} $(\G,\W)$ is a digraph $\G$ paired with a weight function $\W$ that assigns a weight $\W[e]$ to each directed edge $e$ of $\G$. For the sake of generality we let the weight of a directed edge from $\mu$ to $\nu$, denoted $\ew_{\nu\mu} := \W[(\mu\nu)]$, be a $d_\nu$-by-$d_\mu$ complex matrix.
\sjt{We extend the action of $\W$ from edges to walks by defining the weight of a trivial walk $(\mu)$ to be the $d_\mu \times d_\mu$ identity matrix $\I_\mu$, and the weight of a walk $w = \alpha\mu_1\cdots\mu_{\ell-1}\omega$ of length $\ell \ge 1$ to be the right-to-left product of the weights of the edges traversed by $w$: $\W[w] = \W[(\mu_{\ell-1}\omega)]\cdots\W[(\alpha\mu_1)]$. Note that the ordering of the edge weights is suitable for the matrix multiplications to be carried out, the end result being a $d_\omega \times d_\alpha$ matrix. }

\begin{corollary}[Path-sum expression of weighted sums of walks]\label{FactorSumOfWeights}
\sjt{Let $(\G,\W)$ be a weighted digraph, and $\alpha$ and $\omega$ be two vertices on $\G$.
If it exists, the sum of the weights of all walks from $\alpha$ to $\omega$ on $\G$, denoted by $\W\,\Big[\Sigma_{\G;\,\alpha\omega}\Big]=\sum_{w\in W_{\G;\alpha\omega}} \W[w]$, admits a factorised form involving only the weights of prime walks. We term this form a path-sum representation. It is explicitly given by}
\begin{subequations}
\begin{align}
&\W\,\Big[\Sigma_{\G;\,\alpha\omega}\Big]=\sum_{p\in \Pi_{\G;\,\alpha\omega}}\prod_{j=0}^{\ell(p)}\bigg\{\W\,\Big[\Sigma_{\G\backslash\{\alpha,\,\nu_2,\cdots,\,\nu_{j-1}\};\,\nu_{j}\nu_{j}}\Big]\, \ew_{\nu_{j+1}\nu_{j}}\bigg\}\,\W\,\Big[\Sigma_{\G;\,\alpha\alpha}\Big]\,,\\
&\W\,\Big[\Sigma_{\G;\,\alpha\alpha}\Big]=\left(1-\sum_{\gamma\in\Gamma_{\G;\,\alpha\alpha}}\ew_{\mu_{0}\mu_{\ell(\gamma)}}\prod_{j=1}^{\ell(\gamma)}\bigg\{
\W\,\Big[\Sigma_{\G\backslash\{\alpha,\,\mu_2,\cdots,\,\mu_{j-1}\};\,\mu_{j}\mu_{j}}\Big]\, \ew_{\mu_j\mu_{j-1}}\bigg\}\right)^{-1},\label{Saa}
\end{align}
\sjt{where the products are to be constructed right-to-left; $p=\nu_0\nu_1\cdots \nu_{\ell(p)}$ is a simple path of length $\ell(p)$, where we identify $\alpha$ with  $\nu_0$ and $\omega$ with $\nu_{\ell(p)}$ for convenience; and $\gamma=\mu_0\mu_1\cdots \mu_{\ell(\gamma)-1}\mu_0$ is a simple cycle of length $\ell(\gamma)$ from $\alpha\equiv \mu_0$ to itself.}\\
\end{subequations}
\end{corollary}
\begin{proof}
The corollary is an immediate consequence of the formal results of Theorem \ref{FactorSum} on noting that the weight function is \sjt{i) linear, so that $\W\big[w+w'\big] = \W\big[w\big]+\W\big[w'\big]$, and ii) a homomorphism, so that $\W[e_1\conc e_2]=\W[e_2]\W[e_1]$ for any two directed edges $e_1$, $e_2$ such that $e_1\conc e_2$ is non-zero.}

\sjt{The corollary can alternatively be obtained by using} the quiver introduced in the proof of Theorem \ref{FactorSum}. Consider the matrix representation of the mapping $\varphi_{(\alpha)'_{\G}}$. Since $\varphi_{(\alpha)'_{\G}}$ is the inverse mapping of $1_{\alpha}-\varphi_{\Gamma_{\G;\,\alpha}}$, its matrix representation is the matrix inverse of the matrix representation of $1_{\alpha}-\varphi_{\Gamma_{\G;\,\alpha}}$.\qed
\end{proof}
\begin{remark}[Existence of the weighted path-sum]
If $\G$ has finitely many edges and vertices, it sustains only a finite number of primes and the path-sum representation of the weighted series of walks involves only finitely many terms. An immediate consequence is that \emph{the path-sum representation exists even when the sum of walk weights diverges.} In this situation the path-sum has been shown to be the unique analytic continuation of the sum of walk weights, and remains a valid representation of this sum \cite{Giscard2011b}, \cite{Giscard2014b}. This result leads to applications in the field of matrix computations \cite{Giscard2011b}.
\end{remark}

\subsubsection{Extended example: Walks on finite graphs}\label{longexmp}
We now turn to an extended example illustrating the use of Corollary \ref{FactorSumOfWeights}: we obtain the walk generating functions of finite Cayley trees.
For any two vertices $\alpha$ and $\omega$ of a graph $\G$, the walk generating function is an ordinary generating function of the set $W_{\G;\,\alpha\omega}$ defined as \cite{Biggs1993}
\begin{align}
g_{\G;\,\alpha\omega}(z)&:=\sum_{w\in W_{\G;\,\alpha\omega}} z^{\ell(w)}=\sum_n |W_{\G;\,\alpha\omega;n}|\,z^n,
\end{align}
where $|W_{\G;\,\alpha\omega;n}|$ is the number of walks of length $n$ from vertex $\alpha$ to vertex $\omega$ on $\G$. A walk generating function is a weighted sum of walks, with the weight function being simply $\W[e]=z$ for any edge $e$ on $\G$. As a consequence, \sjt{Corollary \ref{FactorSumOfWeights} provides an expression for $g_{\G;\,\alpha\omega}(z)$ that only} involves prime walks.\\


\noindent \textit{Walks on finite path-graphs and cycle graphs}\label{cyclegraph}\\
We begin by determining the prime expression for the walk generating functions of finite path-graphs and cycle-graphs\footnote{Contrary to the generating functions of finite Cayley trees, those of finite path-graphs and cycle-graphs are already known. 
We derive them again to illustrate our results.}. \sjt{Let $\mc{P}_n$ and $\mc{C}_n$  be the ordinary (i.e. undirected) path-graph and cycle graph on $n$ vertices, respectively.} For convenience, we label the vertices of $\mc{P}_n$ from left to right, from $0$ to $n-1$. \sjt{Let $\alpha$ be a vertex of $\mc{P}_n$. Then if $\alpha\neq 0,\,n-1$,
the only simple cycles off $\alpha$ on $\mc{P}_n$ are the two back-tracks $\alpha\to \alpha\pm1\to\alpha$ with weight $z^2$ and, if $\alpha= 0$ or $n-1$, then only one back-track exists.} According to Corollary \ref{FactorSumOfWeights}, the path-sum for $g_{\G;\,\alpha\alpha}(z)$ thus reads
\begin{equation}\label{WGFPn}
g_{\G;\,\alpha\alpha}(z)= \frac{1}{1-z^2 F_{\alpha}(z)-z^2F_{n-\alpha-1}(z)},
\end{equation}
where $F_{\alpha}$ is the continued fraction of depth $\alpha-1$ which represents the weight of the dressed neighbour of $\alpha$,
\begin{equation}\label{ContFracDepthDEF}
F_{\alpha}(z)=\W\big[(\alpha-1)'_{\mc{P}_{n}\backslash\{\alpha\}}\big]=\frac{1}{1-\frac{z^2}{1-\frac{z^2}{\large \hdots}}}=\frac{Q_{\alpha-1}(z)}{Q_\alpha(z)},
\vspace{-1.5mm}
\end{equation}
with $Q_x(u)=\hspace{-.5mm}\, _2F_1\left(\frac{1}{2}-\frac{x}{2},-\frac{x}{2};-x;4u^2\right)$ the Gauss hypergeometric function. Then,
\begin{equation}\label{gaaPn}
g_{\mc{P}_n;\,\alpha\alpha}(z)=\frac{Q_{n-\alpha-1}(z)Q_{\alpha}(z)}{Q_{n}(z)},
\end{equation}
which follows from the identity $Q_{n}(z)=Q_{n-\alpha-1}(z)Q_{\alpha}(z)-z^2Q_{n-\alpha-2}(z)Q_{\alpha}(z)-z^2Q_{n-\alpha-1}(z)Q_{\alpha-1}(z)$.
Now let $\omega$ be another vertex of $\mc{P}_n$. \sjt{Since the graph is symmetric, we may assume without loss of generality that $\omega$ lies to the right of $\alpha$. Since there is only one simple path from $\alpha$ to $\omega$, Corollary \ref{FactorSumOfWeights} yields
\begin{equation}
g_{\mc{P}_n;\,\alpha\omega}(z)=z^{d}g_{\mc{P}_{\alpha-d};00}\cdots g_{\mc{P}_{\alpha-1};00}(z)g_{\mc{P}_n;\,\alpha\alpha}(z),
\end{equation}
where $d = \alpha - \omega \ge 0$ is the distance from $\omega$ to $\alpha$.}
With the result Eq.~(\ref{gaaPn}) we find
\begin{equation}
g_{\mc{P}_n;\,\alpha\omega}(z)=z^d\frac{Q_{n-\alpha-1}(z)Q_{\alpha-d}(z)}{Q_{n}(z)}.
\end{equation}
This gives all the walk generating functions on all finite path-graphs.

We now derive the walk generating functions \sjt{of the cycle graphs $\mc{C}_n$}. For convenience, we label the vertices of $\mc{C}_n$ clockwise from $0$ to $n-1$.
We begin with the walk generating function $g_{\mc{C}_n;\,00}(z)$ for all the cycles off vertex $0$. This is the sum of all cycle weights on a weighted version of $\mc{C}_n$ where all edges have weight $z$. The only simple cycles off $0$ are the two backtracks to its neighbours, \sjt{each of which has weight $z^2$, and two simple cycles of length $n$ (one clockwise and one counter-clockwise) each with weight $z^n$}. Then
\begin{subequations}
\begin{align}
g_{\mc{C}_n;\,00}(z)&=\frac{1}{1-2 z^2\,g_{\mc{P}_{n-1};\,00}(z)-2 z^n\,g_{\mc{P}_{1};\,00}(z)\cdots g_{\mc{P}_{n-2};\,00}(z)g_{\mc{P}_{n-1};\,00}(z)},\label{Cncontfrac}\\
&=\frac{Q_{n-1}(z)}{Q_{n-1}(z)-2 z^2 Q_{n-2}(z)-2 z^n}.\label{HyperAnswer}
\end{align}
\end{subequations}
To obtain Eq.~(\ref{Cncontfrac}), we first used the symmetry of $\mc{C}_n$, noting that $\W\big[(1)'_{\mc{C}_n
\backslash\{0\}}\big]=\W\big[(n)'_{\mc{C}_n
\backslash\{0\}}\big]$ etc. \sjt{Second, we used that} $\mc{C}_n
\backslash\{0\}\equiv\mc{P}_{n-1}$ and similarly, $\mc{C}_n\backslash\{0,1,\cdots,j\}\equiv\mc{P}_{n-j-1}$, $0\leq j\leq n-1$. Then Eq.~(\ref{HyperAnswer}) follows from
Eq.~(\ref{gaaPn}). Now we turn to the walk generating function $g_{\mc{C}_n;\,0d}(z)$ for all walks from 0 to a vertex located at distance $d$, which we assume without \sjt{loss of generality to satisfy} $0\leq d\leq \lfloor n/2\rfloor$. There are two simple paths from $0$ to $d$: one of length $d$ and one of length $n-d$. \sjt{Applying Corollary \ref{FactorSumOfWeights} gives}
\begin{subequations}
\begin{align}
\hspace{-4.5mm}g_{\mc{C}_n;\,0d}(z)&=z^{n-d}g_{\mc{P}_{d};\,00}(z)\cdots g_{\mc{P}_{n-1};\,00}(z)g_{\mc{C}_n;\,00}(z)\\
&\hspace{35mm}+z^dg_{\mc{P}_{n-d};\,00}(z)\cdots g_{\mc{P}_{n-1};\,00}(z)g_{\mc{C}_n;\,00}(z),\nonumber\\
&=\frac{Q_{d-1}(z) z^{n-d}+z^d Q_{n-d-1}(z)}{Q_{n-1}(z)-2 z^2 Q_{n-2}(z)-2 z^n}.
\end{align}
\end{subequations}

%

\begin{figure}[t!]
\begin{center}
\vspace{-3mm}
\includegraphics[width=.9\textwidth]{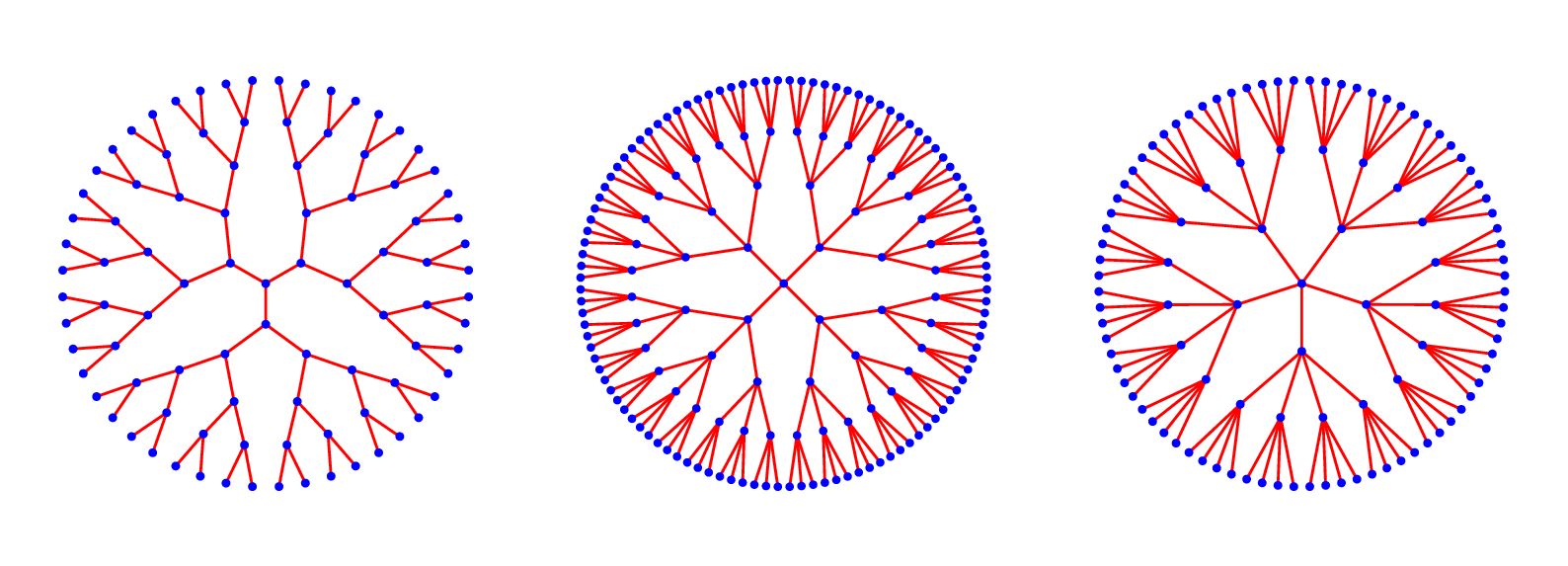}
\vspace{-3mm}
\caption[Illustration of three Cayley trees]{\sjt{Three finite Cayley trees: from left to right $\mc{T}^5_3$, $\mc{T}^4_4$ and $\mc{T}^3_5$.} The corresponding Bethe lattices are infinite in the radial direction.}
\label{Bethe}
\end{center}
\vspace{-3mm}
\end{figure}

\vspace{2mm}\noindent\textit{Walks on finite Cayley trees}\label{FactorExampleBethe}\\
A finite Cayley tree $\mc{T}^\Delta_n$ is an ordinary (i.e. undirected) rooted tree where every vertex within distance $d<\Delta$ from the root $0$ is connected to $n$ other vertices, while vertices at distance $\Delta$ from the root have $n-1$ neighbours \sjt{(see
Fig. \ref{Bethe}).} The quantities $\Delta$ and $n$ are called the radius and bulk connectivity of $\mc{T}^\Delta_n$, respectively.
Finite Cayley trees and their infinite counterparts, the Bethe lattices $\mc{B}_n\equiv \mc{T}_n^\infty$, have found widespread applications in mathematics, physics \sjt{and even} biology \cite{Bethe1935,Baxter1982,Cai1997,Chen1999}.

Even though the finite Cayley tree appears at least as often as the infinite Bethe lattice in applications, the former is usually approximated by the latter which is easier to handle. Indeed, the walk generating functions of the Bethe lattices satisfy \sjt{the following} easily solvable relations\footnote{Called self-consistency relations in the physics literature.}
\begin{subequations}
\begin{eqnarray}\label{GBn}
g_{\mc{B}_n;\,00}(z)&=&\left(1-n\,z^2\,g_{\mc{B}_n\backslash\{0\};\,11}(z)\right)^{-1},\label{GenBeth1}\\
g_{\mc{B}_n\backslash\{0\};\,11}(z)&=&\left(1-(n-1)\,z^2\,g_{\mc{B}_n\backslash\{0\};\,11}(z)\right)^{-1},\label{GenBeth2}
\end{eqnarray}
\end{subequations}
where $0$ and $1$ designate \sjt{an arbitrary vertex, and an arbitrary vertex neighbouring 0, respectively.}
These equations are not fulfilled by finite Cayley trees, which exhibit finite size effects that are often neglected for the sake of simplicity. Yet these effects are generally important due to the large fraction of vertices on the outer-rim of the tree.
In this section we obtain the exact walk generating functions on any finite Cayley tree.

We begin with the walk generating function $g_{\mc{T}^\Delta_n;\,00}$ for the cycles off the root of the tree. There are $n$ backtracks off the root of the tree with weight $z^2$ and therefore
\begin{equation}
g_{\mc{T}^\Delta_n;\,\alpha\alpha}=\frac{1}{1-nz^2F_{\Delta}\big(z\sqrt{n-1}\big)},
\end{equation}
with $F_{\Delta}$ the finite continued fraction of depth $\Delta$ defined in Eq.~(\ref{ContFracDepthDEF}). To see this, observe that each neighbor of the root has itself $n-1$ neighbors on $\mc{T}^\Delta_n\backslash\{0\}$. Thus $F_{\Delta}$ fulfills the recursion relation
\begin{equation}
F_{\Delta}\big(z\sqrt{n-1}\big)=\frac{1}{1-z^2(n-1) F_{\Delta-1}\big(z\sqrt{n-1}\big)},
\end{equation}
with solution $F_{\Delta}\big(z\sqrt{n-1}\big)=Q_{\Delta-1}\big(z\sqrt{n-1}\big)/Q_{\Delta}\big(z\sqrt{n-1}\big)$. The walk generating function is therefore
\begin{equation}
g_{\mc{T}^\Delta_n;\,00}=\frac{Q_{\Delta}}{Q_{\Delta}-nz^2 Q_{\Delta-1}}.
\end{equation}
where the functions $Q_{x}$ are to be evaluated at $z\sqrt{n-1}$. \sjt{We are now in a position to obtain $g_{\mc{T}^\Delta_n;\,0d}$, the walk generating function for walks from the root to a vertex located at distance $d$ from it (where $0\leq d\leq \Delta$).}
Since there is only one simple path from $0$ to $d$, we have
\begin{equation}
g_{\mc{T}^\Delta_n;\,0d}=z^d\,g_{\mc{T}^\Delta_n\backslash\{0,1,\cdots d-1\};\,dd}\times\cdots \times g_{\mc{T}^\Delta_n\backslash\{0\};\,11}\times g_{\mc{T}^\Delta_n;\,00}.
\end{equation}
This simplifies upon noting that the graphs $\mc{T}^\Delta_n\backslash\{0,1\cdots j-1\}$ are truncated Cayley trees of radius $\Delta+1-j$ and with the root connected to only $n-1$ neighbours. \sjt{It follows that} the walk generating functions $g_{\mc{T}^\Delta_n\backslash\{0,1,\cdots j-1\};\,jj}$ \sjt{are equal to} $F_{\Delta+1-j}(z\sqrt{n-1})$ \sjt{for $1 \le j \le \Delta+1$, and we have}
\begin{equation}
g_{\mc{T}^\Delta_n;\,0d}=z^d\frac{Q_{\Delta-d}}{Q_{\Delta}}\frac{Q_{\Delta-1}}{Q_{\Delta-1}-nz^2Q_{\Delta-2}},
\end{equation}
\sjt{where the functions $Q_{x}$ are to be evaluated at $z\sqrt{n-1}$.} In the limit $\Delta\to\infty$, we recover the known results of the Bethe lattice:
\begin{equation}
\lim_{\Delta\to\infty}g_{\mc{T}^\Delta_n;\,0d}=\frac{2^{d+1} (n-1)\, z^d \left(\hspace{-.5mm}\sqrt{1-4 (n-1) z^2}+1\right)^{-d}}{n \sqrt{1-4 (n-1) z^2}+n-2}\equiv g_{\mc{B}_n;\,0d}.
\end{equation}
\sjt{Upon setting $d=0$ we find that $\lim_{\Delta\to\infty}g_{\mc{T}^\Delta_n;\,00}$ fulfills} Eqs.~(\ref{GBn}), as expected.

On $\mathcal{T}_{n}^\Delta$ there are a total of $\tbinom{\Delta+3}{3}-1$ different walk generating functions and we will consequently not derive them all explicitly here. \sjt{However, every one can be derived by applying the result of Corollary \ref{FactorSumOfWeights}.} For example, consider the walk generating function $g_{\mc{T}^\Delta_n;\,dd}$ for a vertex located \sjt{at a distance $d$ from the root, where $0\leq d\leq \Delta$.} We obtain $g_{\mc{T}^\Delta_n;\,dd}$ as the continued fraction of depth $d$
\begin{align}
&\hspace{-3mm}g_{\mc{T}^\Delta_n;\,dd}=\frac{1\,\big|}{\big|1-z^2(n-1)F_{\Delta-d}}-\frac{z^2\,\big|}{\big|1-z^2(n-2)F_{\Delta-(d-1)}}-\\
&\hspace{2.2cm}\frac{z^2\,\big|}{\big|1-z^2(n-2)F_{\Delta-(d-2)}}-\cdots-\frac{z^2\,\big|}{\big|1-z^2(n-2)F_{\Delta-1}-z^2F_{\Delta+1}},\nonumber
\end{align}
where all functions $F_x$ are to be evaluated at $z\sqrt{n-1}$.
In this expression we used the notation of Pringsheim for continued fractions, i.e. $a_0+\frac{a_1|}{|a_2}+\frac{a_3|}{|\cdots}=a_0+\frac{a_1}{a_2+\frac{a_3}{\cdots}}$.

\section{Complexity of the prime factorisation}\label{StarHeightsection}
\sjt{In this last section we present results} concerning the computational complexity of the prime factorisation of walks.

The algorithm for factoring individual walks provided in \S\ref{Algosection} is easily shown to be efficient, with a time complexity for the worst case scenario scaling quadratically with the walk length. \sjt{Conversely, we note that since the primes (i.e.~the simple cycles and simple paths of $\G$)} are difficult to identify, we expect the factorised form of the set of \textit{all} walks $W_{\G;\alpha\omega}$ to be difficult to construct. For example, if \sjt{$\G$ is Hamiltonian,} the Hamiltonian cycle or path must appear in the factorisation of at least one walk set. Consequently, we expect that factoring walk sets requires determining the existence of such a cycle or path, a problem which is known to be NP-complete \cite{Karp1972}.

In order to formalise this observation, we now determine the star-height of the prime factorisation, as given by Theorem \ref{FactorBasis}, of any set $W_{\G;\,\alpha\omega}$. The star-height $h(\mathfrak{E})$ of a regular expression $\mathfrak{E}$ was introduced by Eggan \cite{Eggan1963} as the \sjt{depth of the most deeply-nested Kleene star in $\mathfrak{E}$.} This quantity characterises the structural complexity of formal expressions. \sjt{As Example \ref{exampleK3} illustrates, the prime factorisations of sets of walks typically have a non-zero star-height (see e.g.~\eqref{W11K3Factor}).} Furthermore, the proofs of Theorems \ref{FactorSum} and \ref{FactorSumOfWeights} show that the star-height of $W_{\G;\,\alpha\omega}$ is equal to the depth of the continued fraction generated by Theorems \ref{FactorSum} and \ref{FactorSumOfWeights}. In this section we obtain an exact recursive expression for $h(W_{\G;\,\alpha\omega})$. The following result says that the problem of  evaluating $h(W_{\G;\,\alpha\omega})$ is nonetheless NP-complete on undirected connected graphs:

\begin{theorem}\label{StarHeighGraph}
Let $\G$ be a finite undirected connected graph, possibly with self-loops. Let $\alpha$ and $\omega$ be two vertices on it. Let $\ell_{\alpha}=\max_{\nu\in\mathcal{V}(\G)}\max_{p\in\Pi_{\G;\alpha\nu}}\ell(p)$ be the maximum length of any simple path from $\alpha$ to any other vertex $\nu$ on $\G$. Let $\mathrm{L}\Pi_{\G;\,\alpha}$ be the set of simple paths of length $\ell_\alpha$ starting at $\alpha$.
Then
\begin{equation}
h\big(W_{\G;\,\alpha\omega}\big)=h\big(W_{\G;\,\alpha\alpha}\big)=\begin{cases}\ell_\alpha+1,&\text{if there exists $p\in \mathrm{L}\Pi_{\G;\,\alpha}$ such that the}\\
&\text{last vertex of $p$ sustains a self loop},\\\ell_\alpha,&\textrm{otherwise}.\end{cases}
 \end{equation}
The problem of determining $h(W_{\G;\,\alpha\omega})$ and $h(W_{\G;\,\alpha\alpha})$ is equivalent to determining the existence of a Hamiltonian path starting at $\alpha$. It is therefore NP-complete.
\end{theorem}

\sjt{To prove the Theorem, we begin} by establishing an exact recursive relation yielding the star-height of the prime factorisation of a walk-set. This relation will be necessary to prove Theorem \ref{StarHeighGraph}.
\vspace{1.5mm}\begin{lemma}[Star-height]\label{StarHeight} Let $(\mu_c,\nu_0,\nu_p)\in \mc{V}(\G)^3$. \sjt{Then the star-height of the factorised expression for the set of cycles $W_{\G;\,\mu_c\mu_c}$, denoted by $h\big(W_{\G;\,\mu_c\mu_c}\big)$, is given by the recursive relation
\begin{align}\label{hCycles}
  h\big(W_{\G;\,\mu_c\mu_c}\big) = \begin{cases}0 & \text{if $\,\Gamma_{\G;\,\mu_c}=\emptyset$},
  \\ 1+\underset{\Gamma_{\G;\,\mu_c}}{\max} \:\underset{1 \le i \le c-1}{\max}\:\: h\big(W_{\G\backslash\{\mu_c,\,\mu_1,\cdots,\,\mu_{i-1}\};\,\mu_i\mu_i}\big)
  &\text{otherwise}, \\ \end{cases}
 \end{align}
 where the first maximization in the second line runs over all simple cycles $\mu_c \mu_1\cdots \mu_{c-1} \mu_c\in \Gamma_{\G;\,\mu_c}$.} The star-height $h\big(W_{\G;\,\nu_0\nu_p}\big)$ of the factorised expression for the set of open walks $W_{\G;\,\nu_0\nu_p}$ is
\begin{equation}\label{hOpen}
  h\big(W_{\G;\,\nu_0\nu_p}\big) = \underset{\Pi_{\G;\,\nu_0\nu_p}}{\max} \:\underset{0 \le i \le p}{\max}\:\: h\big(W_{\G\backslash\{\nu_0,\nu_1,\cdots,\nu_{i-1}\};\,\nu_i\nu_i}\big),
\end{equation}
where $(\nu_0 \nu_1\cdots \nu_{p-1}\nu_p)\in \Pi_{\G;\,\nu_0\nu_p}$.
\end{lemma}
\begin{proof}
These results follow from Eqs.~(\ref{WG}, \ref{AG}).
We have $W_{\G;\,\mu_c\mu_c}=C^\ast_{\G;\,\mu_c}$
\sjt{and thus if $C_{\G;\,\mu_c}=\Gamma_{\G;\,\mu_c}=\emptyset$ is empty, then $W_{\G;\,\mu_c\mu_c} = \{(\mu_c)\}$ and $h(W_{\G;\,\mu_c\mu_c})=0$.} Otherwise,
$h(W_{\G;\,\mu_c\mu_c})=1+h(C_{\G;\,\mu_c})$. Now by Eq.~(\ref{AG}) we have \vspace{-1mm}
\begin{equation}
\vspace{-1mm}h\big(C_{\G;\,\mu_c}\big)=\underset{\Gamma_{\G;\,\mu_c}}{\max} \:\underset{1 \le i \le c-1}{\max}\:\: h\big(C_{\G\backslash\{\mu_c,\,\mu_1,\cdots,\,\mu_{i-1}\};\,\mu_i\mu_i}\big),
\end{equation}
and since $W_{\G\backslash\{\mu_c,\,\mu_1,\cdots,\,\mu_{i-1}\};\,\mu_i\mu_i}=
C^\ast_{\G\backslash\{\mu_c,\,\mu_1,\cdots,\,\mu_{i-1}\};\,\mu_i\mu_i}$, Eq.~(\ref{hCycles}) follows. By similar reasoning, Eq.~(\ref{hOpen}) is obtained from Eq.~(\ref{WG}); we omit the details.\qed
\end{proof}

\vspace{1.5mm}\noindent We are now ready to prove Theorem \ref{StarHeighGraph}.
\begin{proof}
We begin by proving the result for $h(W_{\G;\,\alpha\alpha})$.
Let $p_\alpha=(\alpha\nu_2\cdots\nu_{\ell_\alpha})\in\mathrm{L}\Pi_{\G;\,\alpha}$. Consider the cycle $w_\alpha$ off $\alpha$ produced by traversing $p_\alpha$ from start to finish, then traversing the loop $(\nu_{\ell_\alpha}\nu_{\ell_\alpha})$ if it exists, then returning to $\alpha$ along $p_\alpha$.
The proof consists of showing that $w_\alpha$ comprises the longest possible chain of recursively nested simple cycles on $\G$.

To this end, consider the factorisation of $w_\alpha$. Let $L_\alpha$ be equal to $(\nu_{\ell_\alpha}\nu_{\ell_\alpha})$, if this loop exists, or $(\nu_{\ell_\alpha})$, otherwise. Then observe that \sjt{$w_\alpha$ can be written as}
\begin{equation}\label{waFactor}
w_\alpha= b_0\odot \Big(b_1\odot\cdots\odot \big(b_{\ell_\alpha-1}\odot (b_{\ell_\alpha}\odot L_\alpha)\big)...\Big)\,,
\end{equation}
where $b_{0\leq j\leq \ell_\alpha-1}$ is the back-track $b_j=(\nu_{j}\nu_{j+1}\nu_j)\in \Gamma_{\G\backslash\{\alpha,\nu_2\cdots\nu_{j-1}\};\,\nu_j}$, \sjt{and} we have identified $\alpha$ with $\nu_0$ for convenience. Equation (\ref{waFactor}) shows that $w_\alpha$ is a chain of $\ell_\alpha$ (or $\ell_\alpha+1$, if the loop $(\nu_{\ell_\alpha}\nu_{\ell_\alpha})$ exists) recursively nested non-trivial simple cycles, and $W_{\G;\,\alpha\alpha}$ must involve at least this many nested Kleene stars.

To see that this chain is the longest, suppose that there exists a walk $w'$ involving $n>\ell_\alpha$ \big(or $n>\ell_\alpha+1$, if the loop $(\nu_{\ell_\alpha}\nu_{\ell_\alpha})$ exists\big) non-trivial recursively nested simple cycles $c_1,\cdots,\, c_n$; that is, $c_1\odot\big(\cdots \odot(c_{n-1}\odot c_n)\big)\subseteq w'$. Then, by the nestable property, the vertex sequence $s\subseteq w'$ joining the first vertex of $c_1$ to the last internal vertex of $c_n$ defines a simple path $p'$ of length $\ell(p')\geq n>\ell_\alpha$. This is in contradiction to the definition of $\ell_\alpha$, and thus $w'$ does not exist.
Consequently, $h(W_{\G;\,\alpha\alpha})= \ell_\alpha+1$ if the loop $(\nu_{\ell_\alpha}\nu_{\ell_\alpha})$ exists, or $\ell_\alpha$, if there is no self-loop on $\nu_{\ell_\alpha}$.

We now turn to determining $h(W_{\G;\,\alpha\omega})$. Combining Eq.~(\ref{hOpen}) with the result for $h(W_{\G;\,\alpha\alpha})$ obtained above yields
\begin{equation}\label{eq:hstepdemo}
h\big(W_{\G;\,\alpha\omega}\big)= \underset{\Pi_{\G;\,\nu_0\nu_p}}{\max} \:\underset{0 \le i \le p}{\max}\:\:\begin{cases}
\ell_{\nu_i}\big(\G\backslash\{\alpha,\ldots,\nu_{i-1}\}\big)+1&\text{if there is a self-loop on vertex }\nu_{i},\\
\ell_{\nu_i}\big(\G\backslash\{\alpha,\ldots,\nu_{i-1}\}\big)&\mathrm{otherwise},
\end{cases}
\end{equation}
where $\ell_{\nu_i}(\G\backslash\{\alpha,\ldots,\nu_{i-1}\})$ is the length of the longest simple path $p_{\nu_i}$ off vertex $\nu_i$ on $\G\backslash\{\alpha,\ldots,\nu_{i-1}\}$, and $\nu_{\ell_{\nu_i}}$ is the last vertex of $p_{\nu_i}$. Finally, we note that $p_\alpha$ is the longest of all the simple paths $p_{\nu_i}$: since $\G$ is undirected and connected, it is strongly connected, and since $\G\backslash\{\alpha,\ldots,\nu_{i-1}\}$ is a subgraph of $\G$ strictly smaller than $\G$, then $p_{\nu_i}$ must be shorter than $p_{\alpha}$. Therefore Eq.~(\ref{eq:hstepdemo}) yields $h(W_{\G;\,\alpha\omega})= h(W_{\G;\,\alpha\alpha})$.

It follows from these results that in order to determine the star-height of the factorised form of any walk set on an \sjt{undirected connected graph $\G$,} one must determine the existence of a Hamiltonian path on $\G$. Consequently, the problem of determining $h\big(W_{\G;\,\alpha\omega}\big)$ and $h\big(W_{\G;\,\alpha\alpha}\big)$ is NP-complete.\qed
\end{proof}

Theorem \ref{StarHeighGraph} means that just determining the complexity of prime factorisations on ordinary graphs is already quite hard. This result may be considered unsurprising in view of the fact that prime factorisations are known to be difficult to obtain, e.g. in the case of integers. Here, however, the origin of the difficulty is different from that in the case of integer factorisation: it resides in factoring \emph{all} the sets of \emph{all} the walks between any two vertices of a connected graph or in computing the star-heights of the factorised forms.

\section{Summary and Outlook}\label{conc}
In this article we established that walks on any finite digraph $\G$ factorize uniquely into nesting products of prime walks, which are the simple paths and simple cycles on $\G$. We used this result to factorize sets of walks, as well as the characteristic series of all walks between any two vertices of any finite \sjt{(possibly weighted)} digraph, thereby obtaining a universal continued fraction expression for these series. \plg{These results have already found applications in quantum mechanics \cite{Giscard2014}, machine learning \cite{Giscard2014b} and linear algebra \cite{Giscard2011b,Giscard2014a}. Although seemingly disparate, many open questions in these disciplines are unified by their natural formulation in terms of walks. Therefore, the prospect for further applications of the results presented in this article is vast.}

We believe that the unique factorisation property will also find applications in the field of graph characterisation. Indeed, a digraph is, up to an isomorphism, uniquely determined by the set of all walks on it \cite{Lawson2004}. The prime factorisation of walk sets which we provide will reduce the difficulty of comparing walk sets to comparing sets of primes, of which there \sjt{are only} a finite number on any finite digraph.

The factorization of walks into products of simple paths and simple cycles is certainly not the only possible construction of this type on digraphs. In particular, the important points in obtaining resummed expressions for series of walks are the \emph{existence} and \emph{uniqueness} of the factorization of walks into primes. \sjt{Provided these properties are satisfied}, there is a unique way to group walks into families generated by their prime factors. \sjt{We are therefore free to construct different walk factorizations based on different definitions for the walk product, each of which induces a different ensemble of prime walks.} Consequently, as long as
the existence and uniqueness properties hold, we can construct as many representations of walk sets and walk series as there are ways to define a walk product. We will formalize these observations in a future work.

\begin{acknowledgements}
P-L Giscard is supported by Scatcherd European and EPSRC scholarships. \sjt{S.~J.~Thwaite acknowledges support from Balliol College, a Clarendon Scholarship, and the Alexander von Humboldt Foundation.}
\end{acknowledgements}


\bibliographystyle{spbasic}      


\end{document}